\documentclass[a4paper,10pt]{article}
\usepackage{ucs}
\usepackage[utf8]{inputenc}
\usepackage{amsmath}
\usepackage{amsfonts}
\usepackage{amssymb}
\usepackage{amsthm}
\usepackage{rotating}
\usepackage{xcolor}
\usepackage[english]{babel}
\usepackage[T1]{fontenc}
\usepackage{graphicx}
\usepackage{hyperref}
\usepackage{cuted}
\usepackage{verbatim}
\usepackage{subfigure}
\usepackage{multirow}
\usepackage{bbold}
\usepackage{slashed}
\usepackage{indentfirst}
\usepackage{tcolorbox}
\usepackage{placeins}
\usepackage[a4paper, total={7in, 10in}]{geometry}
\usepackage{setspace}

\newcounter{eq}

\newcommand{\bpsi}{\bar{\psi}}

\newcommand{\xl}{\lambda}

\begin{document}


\title{\bf 
Flavor-dependent U(3)  Nambu Jona Lasinio coupling constant
}

\author{ Fabio L. Braghin 
\\
{\normalsize Instituto de F\'\i sica, Federal University of Goias}
\\
{\normalsize Av. Esperan\c ca, s/n,
 74690-900, Goi\^ania, GO, Brazil }}
\date{}

\maketitle

\begin{abstract}
A non-perturbative one gluon exchange quark-antiquark interaction
is considered to compute
flavor dependent U(3)  Nambu-Jona-Lasinio (NJL)-type interaction
of the form $G_{ij, \Gamma} 
(\bar{\psi} \lambda_i \Gamma \psi ) ( \bar{\psi} \lambda_j \Gamma \psi)$
for $i,j=0...8$ and $\Gamma=I, i \gamma_5$
from one loop polarization process with
non degenerate u-d-s quark effective masses.
The resulting NJL-type coupling constants in all channels
are resolved in the long-wavelength limit
and numerical results  are presented for  different choices of an
 effective gluon propagator. 
Leading deviations with respect 
to a flavor symmetric coupling constant
are found 
to be of the order of $(M_{f_2}^*-M_{f_1}^*)^n/(M_{f_2}^*+M_{f_1}^*)^n$, 
for $n=1,2$,
where $M_{f_i}^*$ are the effective masses of quarks $f_1,f_2=u, d$ and $s$.
The scalar channel coupling constants $G_{ij, s}$ can be considerably 
smaller  than pseudoscalar ones.
The effect of the flavor-dependence of  coupling constants for 
 the masses of   pions and kaons
may be nearly of the same order of 
magnitude as the effect of the u,d and s quark mass non-degeneracy.
The effect of these coupling constants 
is also verified  for  some of  the light scalar mesons masses, 
 usually described  by 
quark-antiquark states,
{  and for some observables of the pseudoscalar mesons}
\end{abstract}

\section{Introduction}

Theoretical { investigations} and  predictions for low energy strong interacting systems have 
important
support from QCD effective models among which
the  Nambu-Jona-Lasinio
(NJL) type models \cite{NJL,klevansky,vogl-weise}. 
They are suitable  for describing phenomena related to 
 the dynamical  chiral symmetry breaking (DChSB)
 according to which massive constituent quarks can be defined
and are responsible for most part of  the hadron masses.
The NJL model can be derived in terms of  QCD degrees of freedom 
in different ways \cite{kleinert,PRD2014,chilenos,kondo,coimbra-etc,weise-etal}.
Lately, an effective gluon mass was found to be appropriate to 
parameterize the deep infrared behavior of a gluon propagator 
and quark-NJL coupling constant has been 
identified  roughly  to   $G_{NJL} \propto 1/M^2_G$.
Non-perturbative or effective  gluon propagators   take into account part of the 
non Abelian gluon dynamics 
and they might be suitable
to provide numerical estimates for hadron properties.
Eventually they 
make possible a clear relation of fundamental
processes and fundamental degrees of freedom
  with the NJL model framework and description
since  they are expected, at least,   to produce DChSB.
In addition to  the explicit chiral and flavor
 symmetry breakings due to the non-degenerate current quark mass,
the couplings  to  electromagnetic fields also break these
symmetries
contributing to masses  
\cite{klevansky,review-B,pion-kaon-em,donogue,donoghue89,giusti-etal-latt}
  and coupling constants
 \cite{PRD2016}.
It might be interesting to verify if, and to what extent, NJL coupling constant
 receive flavor-dependent contributions.
This should be of relevance for a
fine tuning description hadron masses and dynamics.

Current quark masses, at the energy scale $\mu =2$ GeV,
are approximately
 $m_u \simeq 2.1$ MeV,
 $m_d \simeq 4.7$ MeV and 
$m_s \simeq 93$ MeV
\cite{PDG}
and 
they get amplified due to the DChSB with the 
formation of quark-antiquark scalar condensates. 
In spite of the need of the electromagnetic corrections to describe fully
hadron  masses
there are well known  strong-interaction contributions, for example,
 for masses of pions and kaons, the quasi-Goldstone bosons, that 
are, respectively,  of the order of 
$m_{\pi^\pm} - m_{\pi^0} \simeq 0.1$ MeV
and $m_{K^\pm} - m_{K^0,\bar{K}^0} \simeq - 5.3$ MeV 
\cite{gasser-leutwyler-82}.
This contribution for the  pions mass difference is very small due to the small  difference 
between up and down quark masses.
The electromagnetic neutral and charged
 pions mass difference is  somewhat larger,  of the order of $4$MeV
\cite{pion-kaon-em,donogue,donoghue89,giusti-etal-latt}.
The charged and neutral kaons mass difference  has the opposite sign of the 
pion mass difference and is larger due to the larger strange quark mass.
Flavor   symmetry  breaking corrections are  small
for light
hadrons 
 but important for a good description of hadrons.
In QCD, 
the quark current masses are the only parameters that control 
flavor symmetry breaking.
 This issue  has far consequences in 
some effective approaches  as in 
Chiral Perturbation Theory,  as an Effective Field Theory for  the
low energy regime \cite{CHPT}, 
By starting from QCD to understand effective models, one might expect that
the flavor symmetry breaking encoded in different quark current masses might
spread and have consequences
 to a variety of parameters and coupling constants in effective models 
by means of 
quantum effects.
Therefore NJL- coupling constants
might also be expected  to receive flavor-dependent  contributions
from quantum effects.
Schwinger-Dyson equations (SDE) approach for light and heavy hadrons 
indicates 
couplings constants might be flavor dependent 
\cite{pattern-flavor-qqint}.
The  lightest scalar mesons, that could be expected to be 
chiral partners of the pseudoscalar ones, 
 do not seem to  be compatible with usual quark model 
due to the apparent impossibility of fitting all the  experimental data with 
quark-antiquark structures as prescribed by the simplest quark model
\cite{pelaez-status,pelaez}
although it might be partially appropriated 
\cite{pelaez-status,PDG,su-etal-npa2007,maiani-etal04}.
One of the specific problems with the attempt to describe some of the 
lightest scalar mesons in a U(3) nonet from the
standard  NJL-model scheme is the inverted 
mass behavior of the $a_0 ( I(J^C) = 1(0^+) )$ and $K^* ( \frac{1}{2}(0^+) )$
\cite{hiller-etal2007}.
In the present work this issue appears again and, although no complete solution 
for this problem is obtained
 or proposed, we expect to show further insights for this problem.
In fact there are several different theoretical calculations
with different proposals  for their structures such as composed by
mixed states with tetraquarks, glueballs, mesons molecules or coupled channels resonances
\cite{pelaez,brigitte-etal,giacosa-etal2018,dmitrasinovic,scalars-1,scalars-2,scalars-2b,scalars-3,WGR-2015}.
The full problem of the  light  scalar mesons structure will not be really addressed in
the present work.
Nevertheless
it becomes interesting to introduce as many 
different effects as possible to test 
their individual contributions and the predictive power of the model.

In this work
 explicit chiral and flavor symmetry breaking contributions to 
the NJL coupling constant are  derived by considering
vacuum polarization  in a
flavor U(3) model in which quark-antiquark interaction is mediated by 
a (non-perturbative) gluon exchange.
These coupling constants
are resolved in the local long-wavelength limit in terms of quark and gluon propagators
and they  will be used to calculate light pseudoscalar and scalar mesons masses.
An effective non-perturbative gluon propagator will be considered such as to
incorporate to some extent
non-Abelian dynamics with the crucial requirement
to  produce dynamical chiral symmetry breaking 
and  the large constituent quark masses  
due to the gluon cloud.
The method extends previous works for u-d-s or u-d degenerate  
quarks \cite{PRD2014,PLB2016}.
Because the logics and steps of the calculation
 has been shown with details in previous works, 
in the next section the main steps  are only briefly  outlined.
In Section 3 numerical estimations for the  long-wavelength local limit for the 
resulting four  point  quark interaction
 are  presented as 
NJL-type flavor-dependent interactions. 
This is done for  two
types of effective
gluon propagators and different values of Lagrangian and effective  quark masses.
 In Section 4 the effects of such flavor-dependent NJL-coupling constants
are verified 
on some light mesons masses.
Because not all the scalars might seemingly be  described by
quark-antiquark states, and the pseudoscalar $\eta-\eta'$ mesons require 
further interactions \cite{note},
the masses of 
 pseudoscalar pions and kaons  and of the 
light scalar $a_0$ and $K^*$ (or $\kappa$)
will be investigated.
{ Being that the flavor-dependent coupling constants $G_{ij}$
were found to describe the light  pseudoscalar mesons masses,
further observables will be presented in Section (\ref{sec:observables})
to assess the change in their values 
if $G_{ij}$ are used to redefine the gap equations.}
In the last Section there is a summary.

\section{  Quark determinant and leading current-current interactions}
 \label{sec:Q-det}

The following low energy
 quark effective action
 \cite{PRC1,holdom,ERV}
  will be considered: 
\begin{eqnarray} \label{Seff}  
Z &=& N \int {\cal D} \left[\bpsi, \psi \right]
\exp \;  i \int_x  \left[
\bar{\psi} \left( i \slashed{\partial} 
- m_f \right) \psi 
- \frac{g^2}{2}\int_y j_{\mu}^b (x) 
{\tilde{R}}^{\mu \nu}_{bc} (x-y) j_{\nu}^{c} (y) 
+ \bpsi J + J^* \psi 
\right] ,
\end{eqnarray}
Where the color  quark current is 
$j^{\mu}_a = \bar{\psi} \lambda_a \gamma^{\mu} \psi$, 
 the sums in color, flavor and Dirac indices are implicit, $\int_x$ 
stands for 
$\int d^4 x$,
$a,b...=1,...(N_c^2-1)$ stand 
for color in the adjoint representation, 
and $m_f$ is the   quark current masses matrix  \cite{PDG}, 
{ 
with indices of the flavor SU(3)  
fundamental representation  $f=u,d$ and $s$.
 The adjoint representation of flavor SU(3) will be used with indices
$i,j,k=0,1..,N_f^2-1$ with the additional matrix $\lambda_0=\sqrt{2/3} I$
to complete the U(3) algebra.}
In several  gauges the gluon kernel { is usually argued to  be }
  written in terms of the transversal and longitudinal components in momentum 
space,
$R_T(k)$ and $R_L(k)$,
as:
$\tilde{R}^{\mu\nu}_{ab} (k) = \delta_{ab} \left[
 \left(g^{\mu\nu} - \frac{k^\mu k^\nu}{k^2}
\right)  R_T (k) 
+  \frac{k^\mu  k^\nu}{ k^2}
R_L (k) \right]$.
{Although  this decomposition may  not be exact in general
because of confinement -related effects
 \cite{lowdon},
it is important to emphasize that numerical results
will be calculated by considering effective gluon propagators
whose contributions from other components might 
be parameterized into these two components.
Besides that, it can be shown that   contributions of  other terms,
of the form $\delta (k^2)$ or derivatives of it,
in the   effective gluon propagator for the results
can be expected to  be
much smaller.}
Even if other terms   arise from the 
non Abelian structure of the gluon sector,
 the quark-quark  
interaction  
   (\ref{Seff})  
is a leading term of the QCD effective action.
The use of a dressed (non-perturbative) gluon propagator 
already takes into account non-Abelian contributions that
garantees
  important effects.
Among these, it will be assumed and required
 that this dressed  gluon propagator provides enough strength for 
DChSB
as obtained for example in
 \cite{cornwall,higa,aoki,SD-rainbow,gluon-prop-sde}.
{To some  extent the present work will follow 
previous developments 
by adopting the 
background field method to introduce the background quarks that,
dressed by gluons, give rise to the constituent quarks.
A complete account of the calculations below was presented 
  in Refs. 
\cite{PRD2014,PLB2016,EPJA2016,PRD2019}.
}

To explore the flavor structure of the interaction in the action
(\ref{Seff}), that will be denoted by $\Omega$,
a Fierz transformation
is performed for the quark-antiquark channel 
resulting in  ${\cal F} (\Omega) = \Omega_F$.
We are only interested in this work in 
 the color-singlet scalar-pseudoscalar states  sector
and vector and axial currents will be  neglected. 
Color non-singlet terms are suppressed
 by a factor $1/N_c$ and they might give rise to higher order 
colorless contributions
\cite{EPJA2016,EPJA2018}.
The following bilocal currents are needed to describe the resulting
terms:
$j_i^q(x,y) =  \bpsi (x) \lambda_i \Gamma^q \psi (y)$ where 
$q=s,p$, for scalar and  pseudoscalar currents:
$\Gamma_{s} = \lambda_i I$ (with  the 4x4  Dirac  identity),
 $\Gamma_{p} = i  \gamma_5 \lambda_i$,
where   $\lambda_i$ are the flavor SU(3) Gell Mann matrices
($i=1...8$)
and   $\lambda_0= I \sqrt{2/3}$.
 The resulting $s$ and $p$ 
non local  interactions   are the following:
\begin{eqnarray}
 \Omega_F =
4 \alpha g^2 
\left\{ 
\left[ j_S^i (x,y)  j_S^i (y,x) + j_P^i(x,y)  j_P^i(y,x)  \right] R  (x-y)
\right\},
\end{eqnarray}
where 
$\alpha =  2/9$ 
and
$ R (x-y) \equiv
 R = 
3 R_T (x-y) + R_L (x-y)$.

Next, the quark field  will be split into  the
 background field, $\psi$ constituent quark,
 and the sea quark field, $\psi_2$, that 
might either  form light mesons 
and the chiral condensate
This sort of decomposition is not exclusive to the 
 background field method (BFM)  
 and it is found in other approaches \cite{PRD2001}.
At the one loop BFM 
 level   it is enough to perform this splitting 
for the  bilinears 
 $\bpsi \Gamma_q \psi$ \cite{BFM,EPJA2018}, and
it  can be written that:
\begin{eqnarray} \label{split-Q} 
\bpsi \Gamma^q \psi &\to& (\bpsi \Gamma^q \psi)_2 
+ (\bpsi \Gamma^q \psi),
\end{eqnarray}
where  $(\bpsi \psi)_2$ 
will be treated in the usual  way as sea  quark of the NJL model
and the full interaction $\Omega_F$ is split accordingly
$
\Omega_F  \to \Omega_1 + \Omega_2 + \Omega_{12}
$
where $\Omega_{12}$ contains the interactions between the two components.
This separation 
   preserves chiral symmetry,
and it might  not be  simply a  low and high energies mode separation.
The auxiliary field method makes possible 
to introduce the light quark-antiquark 
chiral states, the chiral condensate  and mesons.
Therefore this procedure   improves one loop BFM since it 
  allows to incorporate
DCSB. 
Because  it is a standard procedure in the field, it 
will not  be described.
{ 
To make possible a clear evaluation of the effects of the resulting 
NJL-coupling constants
the corresponding gap equations at this level, 
which can arise  for the local limit of a auxiliary scalar field,
will be considered  to be those of the NJL model, given in eq. (\ref{gap-eq})
for the case of coupling constants $G_{ff} = G_0 = 10$ GeV$^{-2}$
as discussed below.
This garantees a clear and fair subsequent comparison of the effects of the flavor
dependent coupling constants.
Otherwise, the relation between effective masses and NJL- coupling constants
would be not clear.
}
The non trivial solutions for these gap equations
allow one to define the quark effective masses,
 $M^*_f = m_f  + \bar{S}_f$ where   $\bar{S}_f$ 
($f=u,d,s$).
The quark kernel can be written as:
\begin{eqnarray} \label{q-prop}
S_{0} (x-y) &=& 
( i \slashed{\partial}  - M^*_f)^{-1} \delta (x-y).
\end{eqnarray}

The aim of this work is to present corrections to the 
NJL-type interaction so that the meson sector in terms of 
auxiliary fields will be neglected.
The
quark determinant can then be written as:
\begin{eqnarray}   \label{exp-1}
S_{d} &=& C_0  
+
\frac{i}{2} \;
Tr \ln
\left\{ 
\left(    
1 + 
{S}_{0} 
  \left(\sum_q {a}_q \Gamma_q j_q \right)
\right)^*
\left( 
1 + 
S_{0}   
 \left(\sum_q {a}_q \Gamma_q j_q \right)
\right)
\right\},
\end{eqnarray}
where 
 the following quantities were  defined:
\begin{eqnarray} \label{Gamma}
 \sum_q  a_q \Gamma_q j_q = \sum_q  a_q \Gamma_q j_q (x,y) = 
 2  K_0   R (x-y) 
 \left[  (\bpsi (y)  \lambda_i  \psi(x))
+ 
 i  \gamma_5 \lambda_i  (\bpsi (y) i \gamma_5  \lambda_i \psi (x)) \right],
\end{eqnarray}
where
$K_0 = \alpha g^2$ and   $C_0 =
\frac{i}{2}  \; Tr \; \ln \left[ {S}_{0}^{ -1} S_{0}^{-1} \right]$
that  reduces to 
a constant in the generating functional.

A  large quark mass expansion is performed
with a zero order derivative expansion
\cite{mosel}
 for the 
local limit.
The first  term of the expansion is a  non-degenerate mass term $M_{3,f}$
proportional to the masses from gap equations
\cite{PRD2019} that will not be investigated further.
The second order terms of the expansion correspond to
 four-fermion interactions  with chiral and flavor symmetries breaking.  
These terms, 
in the local limit,
can be written as:
\begin{eqnarray} \label{Seff-NJL}
{\cal L} &=& 
M_{3,f}   (\bpsi \psi)_f
+
G_{ij, s}  (\bpsi \lambda^i \psi)(\bpsi \lambda^j \psi) + 
G_{ij}
(\bpsi i \gamma_5 \lambda^i \psi)(\bpsi i \gamma_5 \lambda^j \psi)
\nonumber
\\
&=& 
M_{3,f}   (\bpsi \psi)_f
+ 
G_{ij} \left[  (\bpsi \lambda^i \psi)(\bpsi \lambda^j \psi)   +
(\bpsi i \gamma_5 \lambda^i \psi)(\bpsi i \gamma_5 \lambda^j \psi)
\right]
-
G_{ij}^{sb}  (\bpsi \lambda^i \psi)(\bpsi \lambda^j \psi) 
 ,
\end{eqnarray}
where the coefficients were resolved and, 
 after a Wick rotation for the Euclidean momentum space
 in the (very long-wavelength) zero momentum transfer limit,
they  are the following:
\begin{eqnarray} 
\label{M3f}
M_{3,f} &=&   d_1 N_c K_0 \; Tr_D \; Tr_F \; \int 
 \frac{d^4 k }{ (2 \pi)^4} 
S_0 (k) \lambda_i R(k)  ,
\\
\label{G2s}
G_{ij, s} = G_{ij} + G_{ij}^{sb} &=&  
d_2 N_c K_0^2  \;
Tr_D  \; Tr_F
\int
 \frac{d^4 k }{ (2 \pi)^4} 
S_0 (k) \lambda_i R(k) S_0 (k)  \lambda_j R(k)
,
\\
\label{G2}  
G_{ij} &=&  
d_2 N_c K_0^2  \;
Tr_D  \; Tr_F
\int
 \frac{d^4 k }{ (2 \pi)^4} 
S_0 (k) R(k) i \gamma_5\lambda_i  S_0 (k) R(k) i \gamma_5 \lambda_j ,
\end{eqnarray}
where $Tr_D, Tr_F$ are the traces in  Dirac and flavor indices,
$d_n =  \frac{ (-1)^{n} }{2 n}$, 
$S_0(k)$ is  the Fourier transform  of  $S_0(x-y)$.
{
If the effective gluon propagator has other terms
proportional to $\delta (k^2)$, or derivatives, \cite{lowdon}
it can be shown that their resulting contribution will be 
suppressed, if not disappearing,
 with respect to the finite momenta component encoded in $R(k)$,
at least, by factors $1/(16 \pi^4 M_{f_1}^* M_{f_2}^*)$.
Eventually these further contributions may also 
 vanish  because of explicit dependences on 
internal loop  momenta $k_\mu$.
}
{ 
To calculate these traces in flavor indices the following strategy was adopted.
Each of the quark propagators, originally in the fundamental representation,
was written as a combination of kernels in the adjoint representation by 
diagonalizing with the correct diagonal GellMann matrices.
This makes possible an unambigous and straithforward
 calculation of the flavor-traces in these equations.
Quark mass matrix  can be written as:
\begin{eqnarray}
M = M_0 \sqrt{3/2} \lambda_0 + M_3 \lambda_3 + M_8 \sqrt{3} \lambda_8,
\end{eqnarray}
where $M_0, M_3, M_8$ are combinations of the up, down and strange quark
effective  masses.
The quark propagator can then be written as:
\begin{eqnarray}
S_{0m} (k) &=& 
I \left[ A (  i \slashed{k} + M_0) + 2 M_8 C  + \frac{2}{3} M_3 B \right]
+ \xl_3 \left[ B  (  i \slashed{k} + M_0)  + M_3  (A+C) + M_8 B \right]
\nonumber
\\
&&+ 
 \xl_8  \sqrt{3} \left[ C  (  i \slashed{k} + M_0)  + M_8 (A - C)  + 
\frac{M_3 B}{3} \right]  ,
\end{eqnarray} 
where
\begin{eqnarray}
A &=& \frac{1}{3} \left( 
\frac{1}{R_u} + \frac{1}{R_d} + \frac{1}{R_s} \right) ,
\\
B &=& \frac{1}{2}\left( 
\frac{1}{R_u} - \frac{1}{R_d} \right) ,
\\ 
C &=& \frac{1}{6} \left( 
\frac{1}{R_u}+ \frac{1}{R_d} - \frac{2}{R_s} \right) ,
\end{eqnarray}
In these equations $R_f = (k^2 - {M_f^*}^2 + i \epsilon)$.
Flavor traces of up to four Gell Mann matrices were calculated, i.e.
$Tr_F (\lambda_{m}\lambda_i \lambda_n \lambda_j)$ where $m,n=0,3,8$ 
and $i,j=0,...8$, with all combinations.
}
The four-point interactions above (\ref{G2s},\ref{G2}), in the limit of 
degenerate quark masses $m_u=m_d=m_s$, 
 reduce to those of Ref. \cite{PLB2016} 
with a different coefficient due to 
the $U(3)$   group.
This way of writing Eq. (\ref{Seff-NJL}) suggests a  non-unambiguos
definition of a nearly {chiral)  "symmetric part" of   NJL interaction as $G_{ij}$
and the chiral symmetry breaking part $G_{ij}^{sb}$ that arises only
for the scalar sector.
$G_{ij}$ is not really a  chiral-symmetric interaction because of the 
 flavor symmetry breakings for 
all the flavor channels $i,j$.
{ 
The integrals for $G_{ij}$ have two components,
one of them strongly dependent on momentum $k$ and the other
strongly dependent on the quark masses, 
whereas $G_{ij}^{sb}$ is written only in terms of  the second of these integrals.
This difference between the integrals (\ref{G2s},\ref{G2})
 favors the above separation of {\it regular}  
and (strongly)   symmetry
breaking (sb) coupling constant.
An important property of these coupling constants is the following:
\begin{eqnarray}
G_{ij} = G_{ji} , 
\;\;\;\;\;\;\;
G_{ij,s} = G_{ji,s}.
\end{eqnarray}
All the integrals in Eqs.  (\ref{G2s}) and (\ref{G2})
are ultraviolet (UV) finite
and infrared regular if the gluon propagator contains 
a parameter such as a gluon effective mass or Gribov type 
parameter. $G_{ij}, G_{ij}^s$ and  $G_{ij}^{sb}$  have 
 dimension of
mass$^{-2}$.
{ 
For some observables, however,
it is more appropriate to define the
following coupling constants between quark currents in the fundamental representation:
\begin{eqnarray} \label{Gff-Gii}
G_{ij}   (\bpsi \lambda^i \psi)(\bpsi \lambda^j \psi) = 2 G_{f_1f_2}
 (\bpsi \psi)_{f_1} (\bpsi  \psi)_{f_2},
\end{eqnarray}
being that none of the types of  mixing terms, $G_{i\neq j}$ 
or   $G_{f_1 \neq f_2}$, will  be considered in most part of 
 the present work.

}

\section{ Numerical results  }
\label{sec:numerics}

In the following, two types of the  effective gluon propagators,
{
that incorporate the quark-gluon running coupling constant $g$,
are written and  
will be  considered for the numerical
calculations. }
The first effective  gluon propagator
  is  a transversal one
 extracted from 
Schwinger Dyson equations calculations  
\cite{gluon-prop-sde,SD-rainbow}.
 It  can be written as:
\begin{eqnarray} \label{gluon-prop-sde}
 D_{I,2} (k)  = g^2 R_T (k)  &=& 
\frac{8  \pi^2}{\omega^4} De^{-k^2/\omega^2}
+ \frac{8 \pi^2 \gamma_m E(k^2)}{ \ln
 \left[ \tau + ( 1 + k^2/\Lambda^2_{QCD} )^2 
\right]}
,
\end{eqnarray}
where
$\gamma_m=12/(33-2N_f)$, $N_f=4$, $\Lambda_{QCD}=0.234$GeV,
$\tau=e^2-1$, $E(k^2)=[ 1- exp(-k^2/[4m_t^2])/k^2$, $m_t=0.5 GeV$,
$D= 0.55^3/\omega$ (GeV$^2$) and 
$\omega = 0.5$GeV.

The second type of  gluon propagator 
 is based in a longitudinal  effective confining parameterization
\cite{cornwall}
that can be  written  as:
\begin{eqnarray} \label{cornwall}
 D_{II,\alpha=5,6} (k)  = g^2 R_{L,\alpha} (k) &=& 
\frac{ K_F }{(k^2+ M_\alpha^2)^2} ,
\end{eqnarray} 
where 
$K_F = (0.5 \sqrt{2} \pi)^2/0.6$ GeV$^2$, as considered in previous works
\cite{PRD2019,EPJA2018},
with, however, either a constant effective gluon mass 
($ M_5 = 0.8$ GeV) or  a running effective mass given by:
$M_6  =  \frac{0.5}{1 +  k^2/\omega_6^2 }$GeV
for $\omega_6=1$GeV.

It can be noted that these effective gluon propagators exhibit different normalizations
and resulting numerical values  for eqs. (\ref{G2s}) and (\ref{G2}) might be quite different.
{ 
Instead of addressing specific issues on the gluon propagators and their
 normalizations
a more pragmatic approach was adopted so that
the relevant issue is not their normalization but the overall momentum dependence 
that contributes in the momentum integrals that generate the flavor-dependencies of the 
results.
Therefore a normalization procedure, common to all 
effective gluon propagators, to compare different results is needed.
}
In addition to that, to make easier the comparison of the 
flavor-dependent effects  a reference coupling constant
with the value $G_0=10$ GeV$^{-2}$ will be considered and 
 the following normalized quantities will be displayed in the Tables below:
\begin{eqnarray} \label{normaliz} 
G_{ij}^n = \frac{G_{ij}}{ \frac{G_{11}}{10} }, \;\;\;\;\;\;\;
G_{ij,s}^{n} = \frac{G_{ij,s}}{ \frac{G_{11}}{10} } .
\end{eqnarray}
In a first analysis  
the resulting coupling constants, $G_{ij}$ and  $G_{ij,s}$, 
 are expected to be additive corrections to 
a constant NJL-model coupling constant.
Given that  the overall absolute values are not determined,
this sort of multiplicative normalization was chosen
instead of an additive 
correction.
Therefore  coupling constants 
are normalized with respect to the channel
 $G_{ij=11}^n = 10$ GeV$^{-2}$
 that  is close to usual values adopted in the 
literature for the NJL model.
Besides that, this way of normalizing coupling constants 
becomes more appropriate to investigate consequences
for the differences between each of the flavor-channels, in particular 
for the light mesons mass differences.

\subsection{ Flavor-dependence of coupling constants}

Besides the gluon 
propagators, that contain implicitly a
 quark-gluon running coupling constant,  
the quark constituent masses
for the quark propagator are also needed.
Sets of values for the parameters that will be considered 
 are shown in Table (\ref{table:masses}) with the Lagrangian quark masses
$m_u,m_d$ and $m_s$.
These values, together with the ultraviolet (three-dimensional)
cutoff $\Lambda$, were chosen
{ 
such that they satisfy a gap equation with a NJL-coupling constant of reference  $G_0$
and this
 makes possible a comparison of the contribution of the particular flavor -dependent 
coupling constant on the results.
}
Besides that,  it will be required that 
 the resulting neutral pion and kaon masses are in quite good agreement
 with experimental values for the case of  
symmetric coupling constant $G_{ij}= G_0 = 10$ GeV$^{-2}$.
{
A combination of up and down quark masses, with the UV cutoff, determine
neutral (or charged)  pion mass and, with the additional strange quark mass, one obtains the 
neutral (or charged) kaon mass.
}
Although only the effective quark masses are needed  for the estimation
of coupling constants, the other parameters of the Table are needed to
find the mesons masses.
Sets of parameters have two main  labels $X$ and $Y$ which
correspond to different ways of dealing with the scalar mesons channels
below.
{The subscripts in labels $X$ and $Y$ are 
just numbers to identify a set of parameters,
they have no 
physical meaning.
The numerical method used to solve the gap equations and the BSE required to identify each 
set of values for the quark masses and cutoff and for each possible 
solution of the BSE.
 For that, a name $X$ or $Y$ was chosen with numbers 
to identify them. }
$X$ stands for sets of parameters with which 
scalar mesons  masses
 were found by considering the same coupling constants as  
for the pseudoscalar mesons channel,
i.e. $G_{ij} = G_{ij,s}$. 
{ This is equivalent to set  $G_{ij}^{sb}=0$  as discussed above.
 Besides  that, 
 sets of parameters $X$ will produce somewhat
better results, as discussed below.
}
$Y$ stands for sets of parameters with which 
scalar mesons  were found  with 
coupling constants obtained from eqs. (\ref{G2s}) instead of
$G_{ij}$,
i.e. $G_{ij,s} \neq G_{ij}$.

\begin{table}[ht]
\caption{
\small Sets of parameters used in the numerical estimations:
 effective and  Lagrangian  quark masses and ultraviolet  cutoff. 
} 
\centering 
\begin{tabular}{c c c c c c c c } 
\hline\hline 
set  of  & $M_u^*$ &   $M_d^*$ &  $M_s^*$ & 
$m_u$ & $m_d$ & $m_s$ &
$\Lambda$
\\
parameters &  MeV  &       MeV     &    MeV  & MeV   &       MeV     &    MeV  & MeV
 \\
\hline
\\ [0.5ex]
$X_{20}$ =$Y_{18}$ &  389  & 399  & 600 &
3 &  7 &  123 &  675 
\\ [0.5ex]
$X_{21}$ &  392   &  396   & 600 & 
4 & 6 & 123  &  675 
\\ [0.5ex]
$Y_{14}$ & 362 & 362 & 574 & 
5 & 5  & 123 &  665
\\ [0.5ex]
$Y_{19}$  &   391  &  395 &   600  & 
4 &  6 & 163 
& 675
\\[1ex] 
\hline 
\end{tabular}
\label{table:masses} 
\end{table}
\FloatBarrier

In Tables (\ref{table:Gij-Gijs-X}) 
and (\ref{table:Gij-Gijs-Y})
the resulting values of coupling  constants $G_{ij}$
are presented for each set of masses
and for the gluon effective propagators:  $D_{I,2}$,  $D_{II,5}$ and $D_{II,6}$.
In Table (\ref{table:Gij-Gijs-Ys}) the resulting values of $G_{ij,s}$ are presented 
for the same  sets $Y$  of Table (\ref{table:Gij-Gijs-Y}).
The set
with  $G_0$ corresponds to a constant and symmetric choice of reference $G_{ij}=G_0 
\delta_{ij}= 10 \delta_{ij}$ GeV$^2$.
This set, $G_0$, is independent of 
the effective gluon propagator 
and it 
was included to make possible a clearer analysis of the effects of the 
flavor-dependence of the coupling constants on the quark-antiquark mesons masses
as discussed above.
{ Note that $G_0$ has the same value of the normalized $G_{11}^n$ and 
this is important for understanding the role of the flavor-dependent coupling constants
on observables.}
The set of parameters $Y_{18}$ has the same values of $G_{ij}$ 
as the set  $X_{20}$ in Table
(\ref{table:Gij-Gijs-X}) and it was included separated in this Table to 
make simpler the comparison of results, 
in particular for  the scalar mesons channel
 in the next section.
Some entries were not included because they
are equal to  those already displayed  in the Table
{ being CP-conserving interactions}:
\begin{eqnarray}
G_{22} = G_{11}, \;\;\;\;
G_{55} = G_{44}, \;\;\;\;
G_{77} = G_{66}.
\nonumber
\\
G_{22,s} = G_{11,s}, \;\;\;\;
G_{55,s} = G_{44,s}, \;\;\;\;
G_{77,s} = G_{66,s}.
\end{eqnarray}

\begin{table}[ht]
\caption{
\small Numerical results for  $G_{ij}$ for 
the sets of parameters $X$ and different gluon propagators.
The entries for $G_0$ are simply defined in this Table
and they correspond to fixed values independent of
any gluon propagator for the sake of comparisons.
} 
\centering 
\begin{tabular}{c  c c c c c c c c c c } 
\hline\hline 
SET & $G_{11}^n$  & $G_{33}^n$ & $G_{44}^n$ & 
$G_{66}^n$ & $G_{88}^n$ & $G_{00}^n$ &  $G_{03}^n$ & $G_{08}^n$ 
& $G_{38}^n$
\\
& GeV$^{-2}$  &  GeV$^{-2}$ &  GeV$^{-2}$  &  GeV$^{-2}$ & GeV$^{-2}$
& GeV$^{-2}$ & GeV$^{-2}$ & GeV$^{-2}$ & GeV$^{-2}$
\\
\hline
\hline
\\ 
X$_{20}$-$D_{I,2}$  & 10.00 &  10.00 & 9.77  & 9.69 & 7.61  & 8.60  &  0.11  & 1.80  & 0.14   
\\
\hline
\\ 
X$_{20}$-$D_{II,5}$ & 10.00 &  10.00 & 9.82 & 9.76  & 8.13 &  8.90  & 0.08  & 1.41  & 0.10   
\\
\hline
\\ 
X$_{20}$-$D_{II,6}$ &  10.00& 10.00& 9.89  &  9.70 & 7.72  &  8.66 & 0.10 & 1.72  & 0.13   
\\
\hline
\\
X$_{20}$-$G_0$ &  10.00& 10.00&  10.00& 10.00&   10.00& 10.00&  0  & 0  & 0
\\
\hline
\hline
\\
X$_{21}$-$D_{I,2}$  & 10.00 & 10.00 &  9.74 & 9.71  & 7.61 & 8.60 & 0.04 & 1.80  & 0.05 
\\
\hline
\\ 
X$_{21}$-$D_{II,5}$  & 10.00& 10.00& 9.80 & 9.78  &  8.12 & 8.90 & 0.03 & 1.41 & 0.04  
\\
\hline
\\ 
X$_{21}$-$D_{II,6}$ &  10.00& 10.00& 9.76 & 9.73 &  7.72 & 8.66 & 0.04 & 1.72 & 0.05  
\\
\hline
\\ 
X$_{21}$-$G_0$ &  10.00& 10.00&  10.00& 10.00&   10.00& 10.00&  0  & 0  & 0
\\
\hline \hline
\end{tabular}
\label{table:Gij-Gijs-X} 
\end{table}
\FloatBarrier


The interaction channels exclusively of  up and down quarks, 
$G_{11}$ and $G_{33}$, are seen to 
be close to each other, i.e. almost no flavor dependent correction.
In fact, in
the leading order $G_{33} - G_{11} 
\propto (\frac{M_d^*-M_u^*}{M_d^*+M_u^*})^2 \sim 10^{-4}$
that is very  small.
The channels involving strange quarks
 have a larger deviation from the symmetric limit
because   $G_{ij=4,5,8,0} - G_{sym}  
\propto   (\frac{M_s^* - M_d^*}{M_s^* + M_d^*})^n$
for $n=1,2$, where
$G_{sym}$ is obtained by the limit of equal quark masses,
the flavor symmetric limit.
Moreover, larger strange quark masses induce smaller values of the
coupling constants.
This is in qualitative agreement with 
 \cite{pattern-flavor-qqint}.
There are also mixing couplings $G_{i\neq j}$ among   the  neutral  channels
from the diagonal generators of flavor U(3) group, i.e.
for $i,j=0,3,8$.
In the pseudoscalar channel
they are all proportional to the quark mass differences in the leading order:
$G_{03} \propto (M_d^*-M_u^*)/M^*$, 
$G_{08} \propto (M_s^*-M_d^*)/M^*$
 and $G_{38} \propto (M_d^*-M_u^*) (M_s^*-M_d^*)/{M^*}^2$.
Note that the mixing $G_{08}$ has 
 the largest values and the difference between its values 
for the sets of parameters 
$X_{20}$ and $X_{21}$ is too small.
 Together with $G_{03}$ and $G_{38}$
 these effective coupling constants   can be 
associated to the $\eta-\eta'-\pi_0$ mixings.
These mixings, however, will not be investigated in the present work.
For the set of parameters $Y_{14}$ in Table (\ref{table:Gij-Gijs-Y}),  
that contains $m_u=m_d$ ,
it yields $G_{03} = G_{38} = 0$.
The set of parameters with $G_0$ 
necessarily implies  $G_{i \neq j} = 0$.

\begin{table}[ht]
\caption{
\small 
Numerical results for  $G_{ij}$ for 
the sets of parameters $Y$ and different gluon propagators.
The entries for $G_0$ are simply defined in this Table
and they correspond to fixed values independent of
any gluon propagator for the sake of comparisons.
The sets of parameters $Y_{18}$ has the same values of $X_{20}$ in Table
(\ref{table:Gij-Gijs-X}).
} 
\centering 
\begin{tabular}{c  c c c c c c c c c c } 
\hline\hline 
SET & $G_{11}^n$  & $G_{33}^n$ & $G_{44}^n$ & 
$G_{66}^n$ & $G_{88}^n$ & $G_{00}^n$ &  $G_{03}^n$ & $G_{08}^n$ 
& $G_{38}^n$
\\
& GeV$^{-2}$ &  GeV$^{-2}$ & GeV$^{-2}$  & GeV$^{-2}$ &  GeV$^{-2}$
&  GeV$^{-2}$&  GeV$^{-2}$&  GeV$^{-2}$ &  GeV$^{-2}$
\\
\hline
\hline
\\ 
Y$_{14}$-$D_{I,2}$ &  10.00 & 10.00& 9.69 & 9.69 &  7.52 & 8.53 & 0 &  2.39 & 0   
\\
\hline
\\
Y$_{14}$-$D_{II,5}$ & 10.00  & 10.00 & 9.76  & 9.76 &  8.06  & 8.85  & 0  & 1.46  & 0  
\\
\hline
\\ 
Y$_{14}$-$D_{II,6}$
 & 10.00  & 10.00 & 9.71  & 9.71 & 7.63  & 8.60  & 0  & 1.04 & 0
\\
\hline
\\ 
Y$_{14}$-$G_0$
 & 10.00  & 10.00  & 10.00& 10.00& 10.00& 10.00 & 0  & 0  & 0
\\
\hline \hline
\\ 
Y$_{18}$-$D_{I,2}$   &  10.00 & 10.00 & 9.77  & 9.69 & 7.61  & 8.60  & 0.11  & 1.80  & 0.14
\\
\hline
\\ 
Y$_{18}$-$D_{II,5}$ &  10.00 & 10.00 & 9.82  & 9.76  & 
8.13  & 8.90  & 0.08  &  1.41   & 0.10  
\\
\hline
\\ 
Y$_{18}$-$D_{II,6}$ & 10.00  & 10.00 & 9.78  & 9.70
& 7.72  & 8.67  & 0.10  & 1.72  &    0.13
\\
\hline
\\ 
Y$_{18}$-$G_0$  & 10.00  & 10.00  & 10.00& 10.00& 10.00& 10.00 & 0  & 0  & 0
\\
\hline \hline
\\  
Y$_{19}$-$D_{I,2}$  &  10.00 & 10.00 & 9.74  & 9.71 & 7.61  & 8.60  & 0.04  & 2.04 & 0.06 
\\
\hline
\\ 
Y$_{19}$-$D_{II,5}$ & 10.00 & 10.00& 9.80 & 9.78  & 8.12 & 8.90 & 0.03 & 1.33  & 0.04 
\\
\hline
\\ 
Y$_{19}$-$D_{II,6}$ & 10.00& 10.00& 9.76 & 9.73 & 7.72 & 8.66 & 0.04 & 1.95 & 0.05 
\\
\hline
\\ 
Y$_{19}$-$G_0$  & 10.00  & 10.00  & 10.00& 10.00& 10.00& 10.00 & 0  & 0  & 0
\\
\hline \hline
\end{tabular}
\label{table:Gij-Gijs-Y} 
\end{table}
\FloatBarrier

The coupling constants $G_{ij,s}$, eq. (\ref{G2s}) of the scalar sector, 
 Table
(\ref{table:Gij-Gijs-Ys}), are only shown 
for the sets of parameters $Y$ because for the sets of parameters $X$
it was considered that $G_{sb} \to 0$.
The gluon propagator $D_{I,2}$  yields, in average,
considerably  lower values of coupling constants
{ being that the normalization (\ref{normaliz})  is,
at least  in part, 
responsible for that.}
These coupling constants may be  even negative (repulsive),
and the effective gluon propagator $D_{II,5}$ 
yields the largest values. 
{ The relative values 
of the $G_{ij,s}$ with respect to the  $G_{ij}$
are  highly dependent on the quark propagator structure 
similarly to the problems that emerge in form factors 
\cite{PRD2018}.
Nevertheless the relative shift of values of $G_{ij,s}$ 
(and the corresponding
changes in the mesons masses)
 for 
each  of the channels for a particular set of parameters should be meaningful.
}
In average the 
scalar channel coupling constants $G_{88,s}$  and  $G_{00,s}$ are 
somewhat lower than the others.
Detailed investigations of the mesons mixings 
and the whole  scalars mesons octet/nonet  mass problems are outside the scope of this 
work.

\begin{table}[ht]
\caption{
\small Numerical results for the scalar channel  normalized 
coupling constants $G_{ij,s}^n$
with the different sets of parameters $Y$.
} 
\centering 
\begin{tabular}{c  c c c c c c c c c c } 
\hline\hline 
SET & $G_{11,s}^n$  & $G_{33,s}^n$  & 
$G_{44,s}^n$ & $G_{66,s}^n$  & $G_{88,s}^n$ & 
$G_{00,s}^n$  &  $G_{03,s}^n$ & $G_{08,s}^n$ & $G_{38,s}^n$
\\
& GeV$^{-2}$ &  GeV$^{-2}$ & GeV$^{-2}$  & GeV$^{-2}$ &  GeV$^{-2}$
&  GeV$^{-2}$&  GeV$^{-2}$&  GeV$^{-2}$ &  GeV$^{-2}$
\\
\hline
\hline
\\ 
Y$_{14}$-$D_{I,2}$ & 0.31  &  0.31 &  0.11 &  0.11 &  -0.98  & -0.47  & 0 & 0.96  & 0 
\\
\hline
\\
Y$_{14}$-$D_{II,5}$ &  3.03 & 3.03 &  2.87   & 2.87  &  1.25  & 2.00  &  0 & 1.35  & 0  
\\
\hline
\\ 
Y$_{14}$-$D_{II,6}$   & 0.87   & 0.87  & 0.67  &  0.67 &  -0.52 & 0.04 &  0 & 1.04 & 0 
\\
\hline
\\ 
Y$_{14}$-$G_0$  &  10.00  & 10.00 & 10.00  & 10.00  & 10.00  & 10.00 &  0 & 0 & 0 
\\
\hline \hline
\\ 
Y$_{18}$-$D_{I,2}$   &  -0.32  & -0.32  & -0.46 & -0.52 &   -1.37 & -0.95 &  0.07 & 0.79   & 0.09 
\\
\hline
\\ 
Y$_{18}$-$D_{II,5}$  & 2.53   & 2.53  & 2.42  & 2.35  &  0.94 &  1.61 & 0.08  & 1.21 &  0.10
\\
\hline
\\ 
Y$_{18}$-$D_{II,6}$ & 0.28   & 0.29  & 0.15  & 0.09  & -0.88  & -0.41  & 0.07  & 0.87   & 0.10  
\\
\hline
\\ 
Y$_{18}$-$G_0$   &  10.00  &  10.00  &  10.00 &  10.00 &  10.00  &  10.00 &  0 & 0 & 0
\\
\hline \hline
\\  
Y$_{19}$-$D_{I,2}$   &  -0.32  & -0.32  & -0.48  & -0.50  & -1.37  & -0.96  & 0.03  & 0.82  & 0.04  
\\
\hline
\\ 
Y$_{19}$-$D_{II,5}$  & 2.53   & 2.53 & 2.35  & 2.33  &  0.76 &  1.50 & 0.03 & 1.33  & 0.04 
\\
\hline
\\ 
Y$_{19}$-$D_{II,6}$  & 0.28  & 0.28  &  0.09 & 0.06 & -0.95  & -0.47  & 0.03  & 0.92  & 0.04  
\\
\hline
\\ 
Y$_{19}$-$G_0$ &  10.00  &  10.00  &  10.00 &  10.00 &  10.00  &  10.00 &  0 & 0 & 0
\\
\hline \hline
\end{tabular}
\label{table:Gij-Gijs-Ys} 
\end{table}
\FloatBarrier

\subsection{ Effect on some  light mesons masses }

In this section   the effect of the flavor-dependent coupling constants
on the masses of light quark-antiquark pseudoscalar   pions and kaons
is analyzed.
Besides that, the effect on some of the scalar quark-antiquark states,
usually associated to the scalars
 $a_0$ and $K^*$, is also analyzed.
This will be done  according to the following
quark structure \cite{maiani-etal04,hiller-etal2007}:
\begin{eqnarray} \label{scalars-structure}
a_0^0 \sim  (\bar{u} u - \bar{d} d)
, 
\;\;\;\;\;\;
a_0^\pm \sim \bar{u} d , \bar{d} u
,
\;\;\;\;\;
K^*_0, \bar{K}^*_0 \sim  \bar{d} s , \bar{s} d
,
\;\;\;\;\;
K^*_\pm \sim \bar{s} u , \bar{u} s
.
\end{eqnarray}
It  must  be kept  in mind, however, that the scalars sector 
should not be expected to be fully worked out and described due 
to the particularities  of their structures quoted in the Introduction.
{ 
Problems in the description of light scalars  are  particularly
stronger for the $\sigma$  that seemingly cannot be 
a quark-antiquark state as reminded in the Introduction.}

The Bethe-Salpeter equation, or bound state equation, (BSE)
 for the quark-antiquark meson sector in the  NJL model 
is usually investigated  at the  Born approximation level, therefore for 
constant Bethe-Salpeter kernel 
that can be written as $K=2G$ 
\cite{klevansky,vogl-weise,klimt-etal,eguichi-sugawara}
for the case of diagonal interaction $G_{ii}$ therefore by neglecting mixing 
interactions.
It provides the following condition 
to determine a particular  meson mass $P^2_0 = M$ of the channel $\Gamma, i$ 
($\Gamma = I, i\gamma_5$ respectively  for S, PS, scalar or pseudoscalar):  
\begin{eqnarray} \label{pole-bse}
1 - 2 \; G_{ii} \; {I_{\Gamma}^{ii}}_{f_1f_2} (P^2_0=M^2,\vec{P}=0) \; = \; 0,
\end{eqnarray}
for the corresponding flavor  i=0,...8  of the quark-antiquark   state,
when written in terms of the $f_1,f_2=u,d,s$ quark flavors.
{ Note that the dependence of the results on the effective gluon propagator
is encoded in the resulting value of  $G_{ij}$
as discussed above.
}
The following integral, for the  four-momentum $P$ of the  meson, was defined:
\begin{eqnarray} \label{integral-bse}
{I_{\Gamma}^{ij}}_{f_1f_2} (P) = i tr_{F,C,D} \int \frac{d^4 k}{ (2 \pi)^4 }
 \left[ 
S_{0,f_1} (k)  \;
\Gamma \;\lambda_i 
\; 
S_{0,f_2} (k+ P) 
\;  \Gamma \; \lambda_i 
\right],
\end{eqnarray}
where $tr_{F,C,D}$ stands for the traces in flavor, color and Dirac 
indices.
Note that the indices $i,i$ of
the GellMann matrices
 of the adjoint representation  are tied with the indices $f_1,f_2$ of the 
fundamental representation of the quark propagators
{ 
for each particular channel in the  integral and in the coupling constants $G_{ii}$.
In the pseudoscalar channel the following association appears:
for the charged and neutral pions: ($i=1,2$ with $f_1,f_2=\bar{u},d / \bar{d},u$)
and ($i=3$ with $f_1,f_2= \bar{u}u+\bar{d}d$) respectively. 
for the charged and neutral kaons:
($i=4,5$ and $f_1,f_2=\bar{u},s / \bar{s},u$)
and ($i=6,7$ with $f_1,f_2= \bar{d}s/\bar{s}d$) respectively. 
}
{ 
By the usual reduction of eq. (\ref{integral-bse}) with the GAP equation
considered so far that is the one with the coupling constant of reference $G_0$, 
that eliminates the quadratic UV divergence, the following forms  
respectively for the pseudoscalar and scalar mesons are  obtained:
\begin{eqnarray} \label{BSE-Gff-ii}
 (P^2 -  ({M_{f_1}^*}  - {M_{f_2}^*}) ^2 ) G_{ij} I_2^{ij} &=&
\frac{G_{ij}}{2 G_0}  \left( 
\frac{m_{f_1}}{M_{f_1}^* }
+ \frac{m_{f_2}}{M_{f_2}^* }
\right)
+ 1 - \frac{G_{ij}}{G_0} 
,
\nonumber
\\
 (P^2 -  ({M_{f_1}^*}  + {M_{f_2}^*}) ^2 ) G_{ij} I_2^{ij} &=&
\frac{G_{ij}}{2 G_0}  \left( 
\frac{m_{f_1}}{M_{f_1}^* }
+ \frac{m_{f_2}}{M_{f_2}^* }
\right)
+ 1 - \frac{G_{ij}}{G_0} 
,
\end{eqnarray}
When the coupling constants become equal
to $G_0$ and this equation reduces to the 
usual BSE with an unique coupling constant \cite{klevansky,vogl-weise}.
The Goldstone theorem is straightforwardly verified
by considering the usual chiral  limit for which the 
effective quark masses are all equal.
}
{
The  integral $ I_2^{ij}$ in eq.  (\ref{BSE-Gff-ii}) is 
 UV logarithmic divergent and it was 
 solved with the same  3-dimensional UV
cutoff exhibited in Table (\ref{table:masses}).
{ Besides that,  
the  pole of the scalar quark-antiquark  bound state $|P_0|=M_S$,
where $M_S$ is the mass of the scalar meson,
might be located 
in the region of external momenta larger than the
sum of two quark effective masses $P_0 > (M^*_{f_1} +  M^*_{f_2})$
such that
there
might have additional poles in the integrals $I_2^{ij}$
indicating instability of the bound state. 
An IR cutoff   \cite{cutoff-IR},
 $\Lambda_{IR} =  120$ MeV,  was used in this case. 
Its contribution for the pseudoscalar mesons masses can be neglected
as usually is.
{
 The value of this cutoff is
somewhat  smaller than values in the literature
because these
larger values lead to a too  large suppression of the 
momentum integrations modifying  
 the scalar mesons masses and their mass differences.
}
It is  also well known that these integrals might be dependent on the regularization 
method used  \cite{SBK}.
However it has been shown that  the regularization method usually 
does not  modify the light mesons properties preserving
quite well  the 
predictive power of the model
 \cite{kohyama-etal}.
Besides that, if one is interested in the influence of the 
flavor-dependent coupling constants on the energy/mass of the 
quark-antiquark meson bound state, i.e. in mesons mass differences,
the regularization method should not produce 
 leading order effect.
Finally, masslessness of pions and kaons is recovered in the chiral limit
as describe above and in the literature\cite{klevansky}.
}


In Table (\ref{table:masses-X})
results for some of the light mesons masses are presented for the
sets of parameters $X$ and 
 three effective gluon propagators presented above 
($D_{I,2}, D_{II,5}$ and $D_{II,6}$).
The case for value of reference,  $G_0$, which is independent of the 
gluon propagator is also considered.
There are two types of comparisons to be done for a given set $X$ or $Y$ below.
Firstly, by reading the lines of the Tables, one can obtain the
neutral-charged mesons mass difference for the pseudoscalar and scalar mesons.
By reading the  columns of the Tables, always within a particular set of parameter $X$ or 
$Y$,  it is possible to verify the role of the flavor-dependent coupling constants
(for each given effective gluon propagator).
{ Although the masses of neutral mesons were used to choose the particular 
sets of parameters in Table (\ref{table:masses}), the charged mesons masses
are obtained as a consequence of the choice of the sets of parameters, being 
therefore predictions.}
{
It is important to  emphasize again 
that one must be concerned with the mesons mass differences rather than
the absolute values of masses.
This is because slightly 
different absolute values for the masses are easily obtained
whereas the mass difference between neutral and charged
mesons are  consequences of quark masses differences and  also
specific values of $G_{ij}$ or $G_{ij,s}$. 
Furthermore, and more importantly, the 
comparison 
among the results from  different effective gluon propagators ($D_{I},D_{II}$  and $G_0$),
 within a particular set of parameter $X$ or $Y$,
indicates the contribution
of varying specifically the flavor dependent coupling constants
$G_{ij}$ or $G_{ij,s}$.
}
Lower values for the pion masses, of the order of $135$ or $136$ MeV,
 were chosen for the fixed cutoff,
instead of larger values   such as 
$140$MeV
with larger UV cutoffs, 
because an electromagnetic contribution for the charged
 pion mass is also expected.
Besides that, the mass difference between neutral and charged pions
due to strong Interaction is small and it depends, first of all,
 on 
 the up and down  Lagrangian  quark mass difference 
 $\delta_{ud} = (m_d - m_u)$.
Consequently it is  dependent on $M^*_d- M^*_u$,
or, more specifically,   
$(m_{\pi^\pm}^2 - m_{\pi^0}^2)
 \sim (M_d^*-M_u^*)^2/(M^*_u + M_d^*)^2$
as expected \cite{donogue}.
The smaller $\delta_{ud}$ from set $X_{21}$
 induces smaller  neutral and charged pion mass difference.
Besides that, 
the smallness of 
$\delta_{ud}$ also favors smaller charged and neutral kaons mass 
differences.
The neutral-charged pion mass difference for $X_{20}$ is around $0.3-0.4$MeV slightly larger than
the obtained by other sets such as $X_{21}$. 
It indicates that the 
up and down quark masses chosen for this set are slightly larger than needed.
The difference between the coupling constants $G_{11}$ and $G_{33}$, however,
in general 
is very  small (of the order of $10^{-3}$ GeV, not  showed in the Tables above) 
such that, usually, it  does not cause meaningful change  in 
the pion masses. 
The exception was found 
 for the set $X_{20}$ (the one with larger u-d mass difference)
 for which the shifts from the values obtained with $G_0$ can be
of the order of $0.1$MeV, that is slightly smaller than the 
mass difference between neutral and charged pions.

The mass difference between neutral and charged kaons
is also considerably in better agreement with expectations
 for the $X_{21}$ set of parameters than 
for the set $X_{20}$.
These mass differences  are 
proportional to  $(M_{f_1}^*-M_{f_2}^*)/(M^*_{f_1} + M_{f_2}^*)$
(where $f_1,f_2 = u,d $ or $s$)
in agreement with other works 
\cite{donogue}.
However the most interesting comparison 
to be noted  
is the fact that different values for $G_{ii}$,
due to different effective gluon propagators for a given set $X_{20}$ 
or $X_{21}$,
lead to different shifts in the kaons masses.
The coupling constants $G_{44}$ and $G_{66}$ are smaller 
than the constant value $G_0$
for 
all the gluon propagators  and therefore
the kaons masses are shifted to larger values.
These shifts  are around $4-6$ MeV,
although
the resulting effect of each effective gluon propagator considered
might be smaller, of the order 
of $1-5$MeV.
The shifts in kaons masses due to flavor dependent coupling constants
have  nearly the same modulus as 
 the mass difference between neutral and charged kaons.
{ The weak decay constant will be calculated in the next section.}

Concerning the scalar channel:
The leading effect for the scalar mesons masses
are the large quark effective masses.
The conditions  for each of  the scalar quark-antiquark (with flavors $f_1-f_2$)
bound state (\ref{BSE-Gff-ii})  might be written approximately as:
\begin{eqnarray} \label{scalar-masses-app}
M^2_{S,f_1f_2} \sim M_{PS,f_1f_2}^2 
+(M_{f_1} 
+ M_{f_2})^2 + {\cal O} (\frac{1}{\tilde{G}_{ij} 
\tilde{I}_{f_1f_2,S} }),
\end{eqnarray}
where
$M_S$ and $M_{PS}$ are the corresponding scalar and pseudoscalar
mesons masses for that particular channel  (with quark-antiquark $f_1-f_2$),
  $\tilde{G}_{ij}  = G_{ij}$ or $G_{ij,s}$,
and $\tilde{I}_{f_1f_2,S}$
is the UV logarithmic  divergent integral that however depends on $M_{S,f_1f_2}^2$.
This approximate equation is in agreement with particular limits of
equations from other works \cite{dmitrasinovic}.
{ 
The largest contribution for the masses of the scalar mesons masses
come from  the second term, i.e. the sum of the quark and antiquark
effective masses, and also the first term for the $K^*$ with 
quark structure analogous to the kaons.
Besides that, the IR cutoff in the integrals for the scalar mesons masses
makes the integrals to be slightly suppressed with respect to the values
obtained in the pseudoscalar channel.
}
The contributions of the coupling constants of the last term
are usually smaller than
the first two terms.
The resulting $a_0 - K^*$ mass hierarchy, according to the 
structure (\ref{scalars-structure}), is inverted
as it  occurs in the simplest versions
of the  NJL model \cite{hiller-etal2007}.
It is possible to fit correctly  their  masses  either
by  introducing other 
Lagrangian interactions  or  by
seeking specific values of 
quark masses and coupling constants $G_{ij}$ to 
fit the desired  values.
This second procedure,  however,  is
too much artificial  from the physical point of view.
 Therefore the entries in the Tables allow for 
a limited
comparison that may be useful,   mainly, for the dependence 
of the mesons masses on each gluon propagator, or conversely,
on flavor dependent
coupling constants.
It turns out however that it is also an
interesting comparison for the mass
 differences of charged and neutral mesons.
{ The overall pattern of masses  is similar to  the one of the  pseudoscalar channel:
set of parameters $X_{20}$ yields larger neutral-charged mass difference
than the set of parameters $X_{21}$.
However the mass difference is of the order of $2$MeV for the 
$a_0^\pm - a_0^0$ isotriplet and around $4$MeV 
for the isoduplets $K^*_0, K^*_\pm$  for the set of parameters
 $X_{21}$ that is better than the set $X_{20}$ whose mass
difference are quite large.
The shifts in the $K^*$ masses due to the coupling constants
are sizeable only for the set of parameters $X_{20}$.
The experimental values for these neutral-charged mesons 
mass differences are seemingly 
slightly smaller than for 
the case of the kaons for example.
Maybe the flavor dependent coupling constants for the scalar channels
should have the role of compensating the quark-effective mass non degeneracy 
effect on the neutral-charged scalar mesons mass differences.
The effective gluon propagators used in this work  did not provide
the corresponding needed $G_{ij,s}$ to reduce accordingly the
neutral-charged scalar mesons mass differences.
}

\begin{table}[ht]
\caption{
\small  Masses of pseudoscalar and scalar mesons states
for the sets of parameters $X$ and the corresponding 
coupling constants  given above without the electromagnetic mass corrections.
For this set $X$ the bound state equation for scalars and pseudoscalars were 
solved with the same coupling constant $G_{ij}^n$.
In the last line there are  experimental values from \cite{PDG}.
$^*$ it  has an electromagnetic contribution ($\sim 4$MeV).
$^{++}$ comparison valid within the assumption for the scalar mesons
quark flavor structure adopted in (\ref{scalars-structure}).
} 
\centering 
\begin{tabular}{c  c c c c c c c c c c } 
\hline\hline 
SET &   $M_{\pi_0}$  &     $M_{\pi+}$     &   $M_{K0}$  
  &   $M_{K+}$     & $M_{a_0^0}/M_{a_0^\pm}$ 
& $M_{\kappa^0}/M_{\kappa^\pm}$ 
\\
 & 
 MeV  &       MeV    &    MeV 
  &    MeV   &  MeV   &   MeV  
\\
\hline
\hline
\\ 
X$_{20}$-$D_{I,2}$ &  135.8 & 
136.2  & 502  & 492  & 781/789   &  1009/999 & & 
\\
\hline
\\ 
X$_{20}$-$D_{II,5}$ &  
135.9
& 136.2  & 501  & 492  & 780/789  & 1003/995 & & 
\\
\hline
\\ 
X$_{20}$--$D_{II,6}$ &     135.8  & 136.2 &
502  & 492  & 781/789  &  1009/999 &  
\\
\hline
\\
X$_{20}$-$G_0$ &    135.9  & 136.2 &  496 & 488  & 781/789  & 1008/998 & 
\\
\hline \hline
\\ 
X$_{21}$--$D_{I,2}$ & 136.1  & 136.2  &
499  & 495  & 787/789  & 1006/1002 & & 
\\
\hline
\\ 
X$_{21}$--$D_{II,5}$  &   136.1  & 136.2 &
498  & 494  & 787/789  & 1006/1002  & & 
\\
\hline
\\ 
X$_{21}$--$D_{II,6}$ & 
136.1  & 136.2  & 499  & 495  & 787/789    &  1006/1002 & & 
\\
\hline
\\ 
X$_{21}$-$G_0$ &  136.1  & 136.2  & 494  & 490   & 
787/789
 & 1006/1002 & & 
\\
\hline\hline
\\
Exp. & 135 & 139.6$^*$  & 497.6 & 493.7    & 980$^{++}$ &      892-896$^{++}$
\\
\hline \hline
\end{tabular}
\label{table:masses-X} 
\end{table}
\FloatBarrier

%

In Table (\ref{table:masses-Y}) 
the 
masses of pseudoscalar and scalar mesons  are presented
for the sets of parameters $Y_{14}, Y_{18}$ and $Y_{19}$,
with the different effective gluon propagators,
 and the corresponding 
coupling constants  given above in Tables 
(\ref{table:Gij-Gijs-Y}) and (\ref{table:Gij-Gijs-Ys}). 
For these sets $Y$, the bound state equation for scalars and pseudoscalars were 
solved with different  coupling constants,  $G_{ij,s}^n$ and  $G_{ij}^n$ respectively.

In  the pseudoscalar channel, the trends are very similar to the previous table:
results for the set of parameters $Y_{18}$ is the same  as 
those for $X_{20}$ in Table (\ref{table:masses-X})
whereas 
resulting pattern of set of parameters $Y_{19}$ is very similar to the 
one for $X_{21}$. 
Therefore similar conclusions apply.
The effect of the lowering of the coupling constants $G_{44}$ and $G_{66}$
is to push the kaons masses to slightly larger 
values than the masses obtained with $G_0$.
The effect of the  change in the coupling constants on the 
scalar mesons masses is considerably larger  than this effect 
on the pseudoscalar masses. 
And this might be interesting for the 
correct complete description of the scalars structures.
In addition to these sets,  there is a set of parameters $Y_{14}$ for which
$m_u=m_d$ and therefore $M^*_u=M^*_d$ that yield
all pions with the same mass, and all kaons with the same mass as expected.
Nevertheless, the coupling constants $G_{44}$ and $G_{66}$ are slightly, 
but sufficiently, different
for the different gluon propagators $D_{I,2}, D_{II,5}$ and $D_{II,6}$
as much as in the other sets $Y$.

{
The main tendency presented in the scalar mesons masses
is the larger mesons masses. 
This is, in one hand, due to 
contribution of the quark effective masses in
eq. (\ref{scalar-masses-app}).
In another hand,  there is a non leading effect that is the fact that 
 the scalar coupling constants $G_{ij,s}^n$ are pretty much  smaller
than $G_{ij}^n$.
The largest shifts in the scalar masses appear 
in the $K^*$ states because the
relative changes in the
 values of $G_{44,s}-G_{66,s}$
 are larger than changes in  $G_{33,s}-G_{11,s}$
together with strange quark effective mass values.
Besides that,
there is almost  no unique trend for the shifts in the scalar mesons masses.
This is due to the very diverse values obtained for
the scalar channel coupling constants $G_{ij,s}$ 
in Table (\ref{table:Gij-Gijs-Ys}).
The resulting overall mass difference between the $a_0$ and $K^*$, in average,
might be as large as $320$MeV for example for set $Y_{14}-D_{I,2}$ 
or $218$MeV for example for $Y_{14}-D_{II,6}$.
The neutral-charged scalar mesons mass differences
are quite different for each of the set of parameters.
This suggests, again, that both the quark effective mass 
and flavor-dependent coupling constants might contribute 
for a fine-tuning of hadrons spectra and interactions.
The main needed effect of inverting the mass hierarchy of $a_0$ and $K^*$ 
does not occur.

}

\begin{table}[ht]
\caption{
\small 
Masses of pseudoscalar and scalar mesons states
for the sets of parameters $Y$ and the corresponding 
coupling constants  given above without electromagnetic mass corrections.
For this set $Y$ the bound state equation for scalars and pseudoscalars were 
solved respectively with  coupling constants $G_{ij,s}^n$ and  $G_{ij}^n$.
In the last line there are  experimental values from \cite{PDG}.
$^*$ it  has an electromagnetic contribution ($\sim 4$MeV).
$^{++}$ comparison valid within the assumption for the scalar mesons
quark flavor structure adopted in (\ref{scalars-structure}).
} 
\centering 
\begin{tabular}{c  c c c c c c c c c c } 
\hline\hline 
SET 
 &  $M_{\pi_0}$  &     $M_{\pi+}$     &   $M_{K0}$  
  &   $M_{K+}$     & $M_{a_0^0}/M_{a_0^\pm}$ 
& $M_{\kappa^0}/M_{\kappa^\pm}$ 
\\
& 
 MeV  &       MeV    &    MeV 
  &    MeV   &  MeV   &   MeV  
\\
\hline
\hline
\\ 
Y$_{14}$--$D_{I,2}$   &
135.1  & 135.1 & 492  & 492  &  758/758 & 970/970 & 
\\
\hline
\\
Y$_{14}$-$D_{II,5}$    &
135.1  & 135.1 & 491   & 491   & 729/729 & 964/964 & 
\\
 \hline
\\ 
Y$_{14}$--$D_{II,6}$  & 
135.1 & 135.1 &  492  & 492  & 740/740 & 978/978  & 
\\
\hline
\\ 
Y$_{14}$-$G_0$ &  135.1  & 135.1 &  487   & 487   &
725/725  & 946/946 & 
\\
\hline \hline
\\ 
Y$_{18}$--$D_{I,2}$
& 135.8  & 136.2  & 502  & 492  & 824/830  &
1068/1060  & 
\\
\hline
\\ 
Y$_{18}$--$D_{II,5}$ & 135.9 & 136.2  & 
501  & 492  & 788/793   & 1028/1017 & 
\\
\hline
\\ 
Y$_{18}$--$D_{II,6}$   &
135.8  & 136.2  & 502  & 492  & 812/820   & 
1067/1045 & 
\\
\hline
\\ 
Y$_{18}$-$G_0$  &   135.9   & 136.2  & 
496  &  488  &  781/789  &  1009/998 & 
\\
\hline \hline
\\  
Y$_{19}$--$D_{I,2}$    & 136.1 & 136.2 & 
499 & 495  & 831/835  & 1037/1037 & 
\\
\hline
\\ 
Y$_{19}$--$D_{II,5}$ 
& 136.1  & 136.2 &  498 &  494 &  792/793 
 &  1024/1020 & 
\\
\hline
\\ 
Y$_{19}$--$D_{II,6}$  & 136.1  & 136.2  &
499  & 495  &  817/820
& 1048/1045   & 
\\
\hline
\\ 
Y$_{19}$-$G_0$ 
&
136.1 & 136.2 &  494  & 490  & 787/789  &  1006/1002 &  
\\
\hline\hline
\\
Exp. & 135 & 139.6$^*$  & 498 & 494    & 980$^{++}$ &      892-896$^{++}$
\\
\hline \hline
\end{tabular}
\label{table:masses-Y} 
\end{table}
\FloatBarrier

\subsection{
Leading effects on gap equations and other observables
}
\label{sec:observables}

The effects of the flavor dependent coupling constants $G_{ij}$
of Table (\ref{table:Gij-Gijs-X}),  
without the mixing couplings, on different observables are
 presented in this section. 
To have a more complete idea of these effects, observables were
calculated firstly by considering the NJL model with the coupling constant
of reference,
$G_0=10$GeV$^{-2}$, and usual resulting solutions from the gap equations
shown in Table (\ref{table:masses}).
Secondly, by recalculating effective quark masses from gap equations
with  the coupling constants
shown in  Table (\ref{table:Gij-Gijs-X}).

First of all therefore, the new or corrected gap equations
in the Euclidean momentum space  are written as:
\begin{eqnarray}  \label{gap-eq}
{M_f'}^* - m_f 
=
 4 N_c G_{ff} \int \frac{d^4 k}{(2\pi)^4}
 \frac{{{M_f'}^*}}{ k^2 + {{M_f'}^*}^2},
\end{eqnarray}
where $G_{ff}$ were extracted  from eq. (\ref{Gff-Gii}) in the absence of 
both types of mixing interactions:
 $G_{i \neq j}$ and $G_{f_1 \neq f_2}$.
The resulting values  
are presented in the first lines of  Table (\ref{table:observables}) 
for the sets of parameters $X$ and compared to the
initial  values for $G_0$.
{ 
 The sets of parameters $Y$ yield too small values of coupling constants $G_{ij,s}$
which are not strong enough to allow for the DChSB 
from eqs. (\ref{gap-eq}).
These too low values of these coupling constants might be rather a consequence
of the common multiplicative normalization adopted in eq. (\ref{normaliz})
as discussed above.
Therefore, because this normalization have  shown to be extremely appropriated for 
 pseudoscalar   channel as discussed in the last section
and, besides that, the scalar channel is not really completely addressed in this work, 
this discussion will be restricted to the sets of parameters $X$.
 }
The reduced values, mainly  of $G_{00}$ and $G_{88}$ which by the way 
are mostly dependent on the strange quark mass and which 
provide smaller contributions for $G_{ss}$ than for $G_{uu}$ or 
$G_{dd}$,  lead to reduced value of the 
 strange effective quark mass ${M'}^*$.
Since the self consistent way of solving the model presents
some further complications, including instabilities of the solutions,
in the following we present results that indicate the 
tendency of the observables when calculating 
them with flavor -dependent coupling constants and corrected effective masses.
Observables predicted by the model are exhibitted
 in Table (\ref{table:observables}), 
by considering two different ways of calculating them,
compared to 
experimental or expected values (e.v.).

The up, down and strange 
chiral scalar quark-antiquark condensates  are implicitely calculated in
the gap equations and 
they can be written as:
\begin{eqnarray}
< (\bar{q} q)_f > \equiv - Tr  \left( S_{0,f} (k) \right),
\end{eqnarray}
where $S_{0,f} (k)$ is the quark propagator.
In Table (\ref{table:observables}) 
 $< \bar{q} q >_G$ stands for the quark condensates
calculated with flavor-dependent coupling constants $G_{ff}$    
 from Table  (\ref{table:Gij-Gijs-X}),
but with the original quark effective masses $M^*_f$.
Due to the larger reduction of $G_{ss}$ (f=u,d and s)
the strange quark condensate is increased.
Secondly, the condensates calculated with
the  same flavor dependent coupling constants 
but with  corrected effective masses  ${M'_f}^*$
are written as $< \bar{q} q >_{M'}$.
Their values are improved with respect to all the other values
and get closer to lattice calculations.
It is worth stressing, lattice results had been calculated at the energy scale 
of $\mu \sim 1-2$ GeV in the works quoted and others cited therein.

The sea-quark couplings to pions and kaons, $G_{qq,\pi}$ and $G_{qq,K}$,
as the residue of the poles of the BSE at the Born level, eq. (\ref{pole-bse}), 
\cite{klevansky,vogl-weise} are presented as calculated
with the initial quark effective masses and with the  corrected quark masses
by means of the equation:
\begin{eqnarray}
G_{qq PS} =  \left( \frac{ \partial \Pi_{ij} (P^2)
 }{\partial P_0^2 } \right)^{-2}_{(P_0, \vec{P})\equiv 0},
\end{eqnarray}
where the flavor indices are tied with the quantum numbers of the 
meson $PS$ as shown above.
Please note that the sets of parameters $G_0$ do not receive correction from the 
flavor-dependent coupling constants and can be used for   comparison.
The effect of the corrected quark masses is to reduce the difference between
$G_{qq\pi}$ and $G_{qqK}$ because the effective masses $M'_f$ are closer to each other .

The  charged pseudoscalar mesons (pions and kaons) weak decay
 constant were also calculated from \cite{klevansky,vogl-weise}:
\begin{eqnarray}
F_{ps} =  \frac{N_c  \; G_{qq PS}}{4}\; \int \frac{ d^4 q}{(2 \pi)^4}
 Tr_{F,D} \left( \gamma_\mu \gamma_5 
\lambda_i \; S_{f_1} (q + P/2) \lambda_j S_{f_2} (q - P/2) \right),
\end{eqnarray}
where $f_1,f_2$ correspond to the quark/antiquark of the  meson 
and $i,j$ are the associated flavor indices as discussed for eq. (\ref{BSE-Gff-ii}).
In Table (\ref{table:observables}) they are presented for the
flavor dependent coupling constants and corrected masses ${M'_f}^*$.
The values for the  sets of parameters with $G_0$ provide the 
results for  the case of flavor-dependent coupling constants.
Because the strange quark mass decreases with
the use of the flavor-dependent interaction, the kaon decay constant
has its value closer to the pion decay constant being the only 
observable whose behavior is not the  expected one.

Finally, the
pseudoscalar mesons mixing that is responsible for the 
eta-eta' mass difference 
will be shortly addressed according to the following 
ansatz.
The logics of the auxiliary field method was adopted and 
the pseudoscalar flavor-quark current interactions
 can be  exchanged by auxiliary fields 
($P_i \sim \bar{q} i\gamma_5 \lambda_i q$).
Within the auxiliary field method the following 
identification, which yields the correct dimension of each of the  fields,
 can be done by implementing functional delta functions
in  the generating functional \cite{alkofer-etal,osipov-etal}:
$\delta( (\bar{q} \lambda_i q) - \frac{ P_i}{G_0})$.
By considering effective masses for the adjoint representation 
auxiliary fields, $M_{ii}^2 P_i^2$,
 the following  terms with 
mixings can be written for the neutral mesons whose states are obtained from
 the diagonal generators of the algebra:
\begin{eqnarray}
{\cal L}_{mix} &=&
\frac{M_{33}^2}{2} P_3^2 + \frac{M_{88}^2}{2} P_8^2  
+ \frac{M_{00}^2}{2} P_0^2 + 
\frac{G_{08}}{G_0^2}  P_0 P_8 
+
\frac{ G_{03} }{ G_0^2} P_0 P_3 
+ \frac{ G_{38} }{G_0^2}  P_3 P_8 ,
\end{eqnarray}
where $M_{ii}^2$ include the contributions from $G_{i=j}$ derived above.
 The mixing terms $G_{i\neq j}$, however,  are exclusively obtained from the 
one-loop interactions   (\ref{G2}).
The neutral pion  ($P_3$)
 mixings and mass are now neglected and
by performing the usual rotation to
mass eigenstates $\eta, \eta'$  \cite{PDG}
it can be written:
\begin{eqnarray} \label{mix-rot}
| {\eta} > &=& 
\cos \theta_{ps} | P_8 > - \sin \theta_{ps} | P_0 >,
\nonumber
\\
| {\eta}' > &=& 
\sin \theta_{ps} | P_8 > + \cos \theta_{ps} | P_0 >.
\end{eqnarray}
 By calculating it and comparing to the above $0-8$ mixing,
the following $\eta-\eta'$ mixing angle is obtained:
\begin{eqnarray}
 \sin (2 \theta_{ps} ) = 
\frac{ 2 G_{0 8} }{ G_0^2   (M_\eta^2 - M_{\eta'}^2 )  } .
\end{eqnarray}
The values for $\theta_{ps}$ are shown in the  Table
 (\ref{table:observables})
and they are   smaller than the expected values.
As remarked above, the coupling constants $G_{08}$ are 
basically the same for the two different sets of parameters
$X_{20}$ and $X_{21}$.
This new mechanism for mesons mixings may not be
sufficient for describing the full mixing.

}

{

In the last lines of the Table
the reduced chi-square, $\chi^2_{red}$ is shown 
for the  sets of parameters $X$
for which ten  observables have been taken into account, two of which
are fitted parameters/observables ($M_{\pi^0}$ and $M_{K^0}$).
The first $\chi_{red}^2$ was done by using $<\bar{q}q>_G$,
the second by using $<\bar{q}q>_{M'}$ and the third without the 
predictions for the   quark  scalar condensates.
The reason   is that  the values for the chiral  quark condensates
have large (and the largest)
 deviations from the expected values obtained from lattice calculations,
therefore their contributions for the $\chi^2_{red}$ are too large.
So the analysis of the reduced chi square can  be done by
considering separately  the behavior of $<\bar{q}q>$.
Results  show the tendency of $\chi^2_{red}$ when comparing 
 for the initial calculation, for  $G_0$, with the  contribution of the 
flavor-dependent coupling constants by means of the effective masses $M'$.
 Besides that, note that the numerical difference between the 
quark condensates calculated in lattices, LQCD, 
have a large deviation from the NJL-prediction.
Another interesting comparison of the $\chi^2_{red}$, 
that shows specifically  the effects of the flavor -dependent coupling constant
in $\chi_{red, M'}^2$ for a specific set of parameters $X$,
is between the sets of parameters $G_0$ (with no flavor-dependent coupling constants
effects) with the other sets (${I,2}, {II, 5}$ and ${II, 6}$.
The same comparison between $G_0$ and the other sets $X$
 for the reduced chi-squared, $\chi_{red,G}^2$
and $\chi_{red,M'}^2$ (no $<\bar{q}q>$),  might be misleading
because the chiral condensates (whose flavor-dependent corrections are larger) 
either are not corrected by $G_{ij}$ or are not taken into account.

}

\begin{table}[ht]
\caption{
\small
Observables for non self consistent calculations by considering
 the  parameters $X_{20}$ and $X_{21}$ discussed above for 
frozen values of the mesons masses.
(e.v.) correspond to the experimental or expected values. 
{ In the last lines the reduced chi-square are presented 
for three different 
calculations and for  each of the 
sets of parameters, $X_{20}, X_{21}$,  by considering
two fitted observables $M_{\pi^0}$ and $M_{K^0}$.
Masses, decay 
constants  and chiral condensate $< \bar{q}q>^{\frac{1}{3}}$
are written  in MeV, coupling constants, 
$G_{Mqq}$,  $\theta_{ps}$ and $\chi^2$  are dimensionless.}
} 
\centering 
\begin{tabular}{c | c c c c | c c  c c | c}
\hline\hline 
  &    $X_{20}^{I,2}$   &  $X_{20}^{II,5}$  & 
  $X_{20}^{II,6}$ 
& $X_{20}, G_0$ &  
 $X_{21}^{I,2}$ & $X_{21}^{II,5}$
     & $X_{21}^{II,6}$  
 & $X_{21}, G_0$
 & (e.v.)
\\
\hline
${M_u}^* (G_0)$  &     389  & 389  & 389  & 389  &   392 & 392  & 392  & 392 
\\
${M'_u}^* (G_{uu})$ &   307  & 325  &  311 &  389   &    310  & 328  & 314  & 392  
\\
${M_d}^* (G_0)$ &  399  & 399  & 399  & 399  &   396  &  396  
& 396 &  396
\\
${M'_d}^* (G_{dd})$ &  319   & 336  & 325 & 399  &   316   & 333  & 320  & 396 
\\
${M'_s}^* (G_{0})$ &   600  & 600 &  600 &  600   &  600 & 600 & 600  & 600 
\\
${M'_s}^* (G_{ss})$ &    349   & 400  & 360  & 600    & 349  &  400 &  360 &  600
\\
\hline
$- < \bar{u} u >_G^{\frac{1}{3}}$  &       348 &    346  &  348 &   338  &    349 & 347
&   348 &  338
 & 240-260 \cite{jamin,flag}
\\ [0.5ex]
$- < \bar{u} u >_{M'}^{\frac{1}{3}}$   &  322 & 326 &  322 &  338  & 322 & 326
& 323
 &
 338 &
\\ [0.5ex]
$- < \bar{d} d >_G^{\frac{1}{3}}$    &    350 &   348 & 349 & 340   & 
  350 & 347
& 
349  &  340 
 &  240-260 \cite{jamin,flag}
\\ [0.5ex]
$- < \bar{d} d >_{M'}^{\frac{1}{3}}$    &   324 &   328  & 325 &  340   &  
  324 & 328
&  325 &  340 
\\ [0.5ex]
$- < \bar{s} s >_G^{\frac{1}{3}}$   &  424  &  407 & 420 &  363
 &    424 & 407
& 
420 &   363
 & 290-300 \cite{davies-etal}
\\ [0.5ex]
$- < \bar{s} s >_{M'}^{\frac{1}{3}}$   &   331 &  340 & 333 &   363 
 &      331 & 340
&  333   &   363
\\
\hline
$G_{qq \pi} (M')$   &  3.28 &   3.40 &  3.31 &  3.83 
 &  
 3.28 & 3.40 & 3.31
&   3.83 & 
\\
$G_{qq K}  (M')$   &   3.38 & 3.61 &  3.43 & 4.59 
 &   3.39  & 3.63
&  3.44 & 4.59
\\
\hline
$f_\pi (M')$  &    95.6 & 97.5     & 96.0 &   103.1 &  
 95.6 & 97.6 & 96.0
&  103.1 
&     92 MeV 
\\
$f_{K} (M')$  &   97.8   & 100.8 &  98.5 &  107.3
 &  97.6 &  100.7 & 98.3
 & 107.2
 &   111 MeV
\\
\hline
$\theta_{ps}^{(0,8)} (0)$
 & 
  -8.2$^{\circ}$
  &   -6.5$^{\circ}$ & -7.9$^{\circ}$  &  0.0
 &     -8.3$^{\circ}$ & -6.5$^{\circ}$ & -7.9$^{\circ}$ & 0.0
&   $- 11^{\circ}/-24^\circ$ \cite{PDG}
\\
\hline\hline
 $\chi^2_{red, G}$   &  313 & 295 & 309 & 244 
 &  314   &  295  & 309  & 245 
\\
\hline
 $\chi^2_{red, M'}$
 &  87  & 92  & 88  & 116 
 & 86  & 92  & 87 &  117
\\
\hline
 $\chi^2_{red, M'}$ (no $<\bar{q}q>$)
 &  23 & 22 & 22 & 19 & 29 & 27  & 29 & 24 & 
\\[1ex] 
\hline 
\end{tabular}
\label{table:observables} 
\end{table}
\FloatBarrier

\section{ Final remarks }

In this work,  flavor symmetry breaking  corrections to the NJL-type quark 
interactions  were derived from a 
quark-antiquark interaction mediated by dressed gluon exchange.
All resulting coupling constants 
are directly proportional to the quark-gluon running coupling constant
and they 
depend   on the quark and  (effective) gluon propagators.
Whereas the coupling constants $G_{ij}$,  defined
 as almost chiral-symmetric couplings,
 can be associated to  the pseudoscalar channel, 
the coupling constants $G_{ij, s}$ of the scalar channels 
can be  numerically  quite smaller and they  
present stronger dependence  on the gluon propagator.
Different sets of coupling constants $G_{ij}$ and $G_{ij,s}$ 
were obtained from sets of quark masses 
 by employing  different effective gluon propagators.
Different sets of parameters
that  were labeled as   $X$ or $Y$
{ correspond to different solutions of a NJL gap equation for a
coupling constant $G_0=10$GeV$^{-2}$ of  reference. }
Although the quark effective masses differences  
induce the flavor-dependence of coupling constants, $G_{ij},G_{ij,s}$,
the effective gluon propagator also slightly   contributes   for the determination
of their relative strength.
The effects of the  flavor-dependent coupling constants
were identified by comparing  results obtained with them
 with the results
for the fixed reference value $G_0$ in the Tables.
{ To  make possible a correct assessment of the 
effects of the flavor-dependent coupling constants,
a normalization for $G_{ij}$ was proposed being 
defined to the  pseudoscalar channel and therefore more reasonable
for pseudoscalar  interactions. 
}
The channels with strangeness develop smaller  values of $G_{ij}$,
i.e. larger deviations from $G_0$, due to the larger strange quark mass.
The mixing type interactions $G_{i\neq j}$ and $G_{i\neq j,s}$ 
were found to be,
in averaged, small,
being proportional to the quark effective mass
differences: $M^*_s-M^*_u$ and/or $M^*_d-M^*_u$.
One set of parameters, $Y_{14}$, was defined with $m_u=m_d$
and the resulting coupling constants $G_{ij}$ and mesons masses calculated
with it, carry this information:
$G_{0,3}=G_{38}=0$ and also $m_{\pi^0}=m_{\pi^\pm}$
and so on. 
{ These mixings, $G_{i\neq j}$, yield light mesons mixings and,
although the mixing angle for  
 the $\eta-\eta'$ mixing has been calculated,
other consequences
  will be investigated  in another work.
}

The charged and neutral
 pion mass difference  was found to be very small and of the order of 
$0.1$ MeV and it is  basically due to the small
up -down quark mass difference, in agreement with expectations.
The effect of the coupling constants $G_{11},G_{33}$ 
is however still slightly  smaller and it was  almost  not really identified,
except for a particular set of parameter with slightly 
 larger u-d quark mass difference.
The remaining part of the pion mass difference comes from
electromagnetic effects that were not calculated in this work.
The neutral-charged kaons mass difference
was obtained to be of the order of 
$4-10$MeV.
Both the quark mass difference and the flavor dependent couplings
however yield  kaon mass differences of the same order of magnitude.
The  flavor dependent coupling constants $G_{44}$ and 
$G_{66}$ induce mass shifts of the order of 2-4 MeV but it could reach
 6 MeV for some of the sets of  parameters.
Both flavor dependences should have to be considered in the NJL:
the mass and coupling constant flavor dependence.
This goes along the very idea of 
considering the NJL as an effective model for QCD, being that 
the initial QCD-flavor dependence, parameterized in the quark Lagrangian masses,
would have consequences for all the effective  parameters of the resulting 
effective model.
This is 
analogous to the flavor-breaking dependence of parameters in EFT, such 
as ChPT.
{

}

 The effects of the flavor-dependent coupling constants  on some of 
the  light scalar  mesons, $a_0$ and  $K^*$ (or $\kappa$),
follow nearly the same patterns of the pseudoscalar mesons.
The shifts of the masses due to changes in the coupling constants, however,
might   not be as large as the changes in the 
quark effective masses.
The
 largest  effects due to varying $G_{ij,s}$ were  found to be of the order of 30-50 MeV.
The usual problem of inverted hierarchy of the scalar mesons
$a_0$ and $K^*$ showed up because of the pattern
of the values of coupling constants does not correct it.
Other contributions for these scalar channel quark-antiquark
interactions are  expected  to correct this inverted mass hierachy
\cite{dmitrasinovic,hiller-etal2007}.
Nevertheless, the $K^*(890)$ mesons masses
might be approximately obtained by
the following ad hoc set of parameters:
$m_u=3$ MeV, $m_d=5$MeV, $m_s=143$MeV,
$\Lambda = 840 $MeV,
 $\Lambda_{IR}=120$MeV, 
$G_{44}=3.65$ and $G_{66}=3.60$,
which yields values close to the experimental ones:
$m_{K^*_0} = 901$MeV and $m_{K^*_\pm} = 888$ MeV.
The same type of fitting is possible for  the $a_0(980)$ mesons
although the physical meaning  or content is not clear.
The experimental values for these neutral-charged 
mesons mass differences, however,  might be   smaller than for 
the case of the  kaons for example.
The effective gluon propagators used in this work however 
can provide
the corresponding needed  $G_{ij}$ or  $G_{ij,s}$ to reduce accordingly the
neutral-charged scalar mesons mass differences.
However, one might expect that  both 
non degeneracy of quark   masses values 
and flavor-dependent coupling constants contribute to
keep the correct experimental behavior of 
neutral-charge scalar mass differences.

{
The resulting coupling constants found above 
define new gap equations as presented in the last section.
{ 
The corrected effective mass provide observables, as calculated in
Section III.C, in average in better agreement
 with expected or experimental values.
To conclude that, note
that the  cutoff and current quark masses were  kept fixed in such a way 
to show clearly the effects of the flavor-dependent coupling constants.
The main source of shifts of the values is the 
strange quark effective mass that decreases with the reduction
of the coupling constant $G_{ss}$.
The consequences in the kaon decay constant and on the strange quark 
condensate are clear.
An interesting issue to note is the new mechanism for 
the mesons mixings by means of the 
resulting coupling constants $G_{i\neq j}, G_{i \neq j, s}$ 
or equivalently $G_{f_1 \neq f_2}$.
The  values found however 
 were not enough to reproduce the complete
$\eta-\eta'$ mass difference.
Further calculations are needed 
and they should help to constraint further 
the corresponding components of the (effective) 
quark-antiquark interactions.
{
The reduced chi-squared was calculated for 
ten  observables being  two  of them fitted observables.
The sets of parameters with the flavor-dependent coupling constants
were shown to provide considerably better results.
}
{
The kaon decay constant is the only observable that 
present worse values when
receiveing corrections due to $G_{ij}$, in the present calculation.
On the other hand, the  values of 
scalar chiral condensates are largely improved.
}
}
These solutions for the  corrected  gap equations induce  further ambiguities to 
define either a new cutoff or different values for Lagrangian quark masses.
As such, a fully self consistent numerical 
calculation may be  expected for which the gap equations and the 
flavor-dependent coupling constants are solved at once. 
In this program 
 non stable results easily appear
since shifts in quark effective masses 
might be reasonably large for fixed cutoff and current quark masses.
{
This problem may be
 worsen if weaker NJL coupling constants $G_0 < 10$ GeV$^{-2}$
are considered.}
This problem prevents a direct and immediate
 self consistent solution for the gap equations, coupling constants
 and eventually BSE
described above.
A more complete account of the mixing interactions
 contributions  
for the light mesons spectra and other observables
 will be treated separately in another work.
}

 \vspace{0.5cm}

\centerline{\bf Acknowledgements}

F.L.B. is member of
INCT-FNA,  Proc. 464898/2014-5
and  he acknowledges partial support from 
CNPq-312072/2018-0 
and  
CNPq-421480/2018-1.
The author thanks a  short conversation with B. El Bennich 
  another one with G. Krein.


\begin{thebibliography}{00}




\bibitem{NJL}
Y. Nambu, G. Jona-Lasinio, 
Dynamical Model of Elementary Particles Based on an Analogy with Superconductivity I,
Phys. Rev. {\bf 122},  345  (1961).

\bibitem{klevansky}
S. P. Klevansky 
The Nambu-Jona-Lasinio model of quantum chromodynamics,
 Rev. Mod. Phys. 64,    649 (1992).


\bibitem{vogl-weise}
U. Vogl, W. Weise,    
{The Nambu and Jona-Lasinio model: Its implications for Hadrons and Nuclei},
 Progr. in Part. and Nucl. Phys. {\bf 27},   195 (1991).

\bibitem{kleinert}
H. Kleinert,   1978
{\it in Erice Summer Institute 1976,
Understanding the Fundamental Constituents of Matter}, 289,
Plenum Press, New York, ed. by A. Zichichi.

\bibitem{PRD2014}
A. Paulo Jr., F.L. Braghin,  
Vacuum polarization corrections to low energy quark effective couplings,
Phys. Rev. {\bf D 90},  014049  (2014).

\bibitem{chilenos}
J. L. Cortés, J. Gamboa, L. Velásquez,  
A Nambu-Jona-Lasinio like model from QCD at low energies,
Phys. Lett. B 432,   397 (1998).

\bibitem{kondo} K.-I. Kondo,
Toward a first-principle derivation of confinement and 
chiral-symmetry-breaking crossover transitions in QCD,
 Phys. Rev. D 82,   065024 (2010).


\bibitem{coimbra-etc}
P. Costa, O. Oliveira, P.J.Silva, 
 What does low energy physics tell us about the zero
momentum gluon propagator, 
Phys. Lett. B 695,  454 (2011).


\bibitem{weise-etal}
T. Hell, S. Rossner, M. Cristoforetti, W. Weise, 
Dynamics and thermodynamics of a nonlocal Polyakov--Nambu--Jona-Lasinio model with running coupling, 
Phys. Rev. D 79,  014022 (2009).

\bibitem{review-B}
V.A. Miransky I.A. Shovkovy,
Quantum field theory in a magnetic field: From quantum
chromodynamics to graphene and Dirac semimetals,
Phys. Rept. 576,  209 (2015).
 J. O. Andersen, W. R. Naylor, A. Tranberg, 
Phase diagram of QCD in a magnetic field 
 Rev. Mod. Phys. {\bf 88},   025001 (2016).


\bibitem{pion-kaon-em}
S. Basak {\it et al}, 
Lattice computation of the electromagnetic contributions
to kaon and pion masses,
Phys. Rev. D99, 034503 (2019).

\bibitem{donogue}
J. F. Donoghue, A.F. Perez,
The Electromagnetic Mass Differences of Pions and Kaons,
Phys.Rev. D55, 7075 (1997).

\bibitem{donoghue89}
J.F. Donoghue, Light quark masses and chiral symmetry,
Annu. Rev. Nucl. Part. Sci. 39, 1 (1989).


\bibitem{giusti-etal-latt}
D. Giusti, {\it et al},
Leading isospin-breaking corrections to pion, kaon 
and charmed-meson masses with Twisted-Mass fermions
Phys. Rev. D95, 114505 (2017).


\bibitem{PRD2016}
F. L. Braghin, 
SU(2) low energy quark effective couplings in weak external magnetic field,
Phys. Rev. {\bf D 94}, (2016) 074030.



\bibitem{PDG}
M. Tanabashi et al. (Particle Data Group), 
 Phys. Rev. D 98,   (2018) 030001.

\bibitem{gasser-leutwyler-82}
J. Gasser, H. Leutwyler, 
Quark masses,
Phys. Rept. 87, 77 (1982).




\bibitem{CHPT}
D.B. Kaplan, Lectures on effective field theory - ICTP-SAIFR, (2016).


\bibitem{pattern-flavor-qqint}
M. Chen, L. Chang, 
A Pattern for the Flavor Dependent Quark-antiquark Interaction,
Chin. Phys C 43,  (2019)  114103.
M. Chen,
Lei Chang, 
Y.-x. Liu,  
Bc Meson Spectrum Via Dyson-Schwinger Equation and Bethe-Salpeter Equation
Approach, 	Phys. Rev. D 101, (2020) 056002


\bibitem{pelaez-status}
J.R. Pelaez,
Status of light scalar mesons as non-ordinary mesons,
 Journ. of Phys.: Conference Series 562, 012012 (2014).


\bibitem{pelaez}
 J. R. Pelaez,
From controversy to precision on the sigma meson: a review on the status of
 the non-ordinary $f_0(500)$ resonance, 
 Phys. Rep. 658, 1 (2016).


\bibitem{su-etal-npa2007}
M.X. Su, L.Y. Xiao, H.Q. Zheng, 
On the scalar nonet in the extended
Nambu–Jona-Lasinio model,
Nuclear Physics A 792, 288 (2007).

\bibitem{maiani-etal04} 
L. Maiani, F. Piccinini, A.D. Polosa,
V. Riquer, 
New Look at Scalar Mesons,
 Phys. Rev. Lett. 93, 212002 (2004)


\bibitem{hiller-etal2007} 
A.A. Osipov, B. Hiller, A.H. Blin, J. da Provid\^encia,
Effects of eight-quark interactions on the hadronic
vacuum and mass spectra of light mesons,
Annals of Phys. 322, 2021
(2007).


\bibitem{brigitte-etal}
 A. A. Osipov, B. Hiller, A. H. Blin, J. da Providencia,
Effects of eight-quark interactions on the hadronic vacuum and mass spectra of light mesons,
Annals of Phys. 322, 2021 (2007).


\bibitem{giacosa-etal2018}
F. Giacosa, 
A. Koenigstein,  R.D. Pisarski,
How the axial anomaly controls flavor mixing among mesons,
Phys. Rev. D97, 091901(R) (2018).

\bibitem{dmitrasinovic} V. Dmitrasinovic, 
UA1 breaking and scala mesons in 
the Nambu and Jona-Lasinio model
Phys. Rev. C 53, 1383 (1996).




\bibitem{scalars-1}
Some proposals not fully  confirmed:
M. Albaladejo,  J. A. Oller,  Identification of a Scalar Glueball,
Phys. Rev. Lett. 101, 252002 (2008).
C. Amsler,  F. E. Close, Evidence for a scalar glueball,
Phys. Lett. B 353, 385 (1995).

\bibitem{scalars-2} E. Klempt and A. Zaitsev, 
Glueballs, Hybrids, Multiquarks. Experimental facts versus QCD inspired concepts,
Phys. Rep. 454, 1 (2007).

\bibitem{scalars-2b}
A. H. Fariborz, R. Jora, J. Schechter, Toy model for two chiral nonets,
Phys. Rev. D 72, 034001 (2005).

\bibitem{scalars-3}
Long-Cheng Gu {\it et al}, 
Scalar glueball in radiative $J/\psi$ decay on the lattice,
Phys. Rev. Lett. 110, 021601 (2013).
\\
F. Brünner, D. Parganlija, A. Rebhan, 
Glueball decay rates in the Witten-Sakai-Sugimoto model, 
Phys. Rev. D 91, 106002 (2015); 
Erratum Phys. Rev. D 93, 109903 (2016)


\bibitem{WGR-2015}
T. Wolkanowski, F. Giacosa, D.H. Rischke, 
$a_0(980)$ revisited, 
Phys. Rev. D 93, 014002 (2016).



\bibitem{PLB2016}
F.L. Braghin, 
SU(2) Higher-order effective quark interactions from polarization,
Phys. Lett. {\bf B 761}  (2016)  424.


\bibitem{note} The light pseudoscalar mesons mass hierarchy is fully  
 described by a variety of approaches when introducing degrees of freedom 
usually related to the U$_A$(1) symmetry breakdown \cite{UA1}.
This last term is effectively  responsible for the so called $\eta-\eta'$ mixing
that makes the masses of the full pseudoscalar nonet to pin down experimental values
\cite{thooft-mesons,klevansky,vogl-weise,PDG}. 
Interestingly it has been pointed out in \cite{PRD2014} that
the sixth order quark interaction in flavor U(3) NJL model, usually
pointed out as consequence of a U$_A$(1)  symmetry breaking phenomenon,
 can also be obtained 
from polarization corrections of NJL-type models.

\bibitem{UA1}
E. Witten, Nucl. Phys. B 156, 269 (1979); G. Veneziano, Nucl. Phys. B
159, 213 (1979).
G. ’t Hooft, Phys. Rev. D 14,  3432 (1976); Erratum: ibid D 18, 2199  (1978).

\bibitem{thooft-mesons} C. Rosenzweig, J. Schechter and C. G. Trahern, Phys. Rev. D 21,
3388 (1980).
 R. Alkofer and I. Zahed, Mod. Phys. Lett. A 4, 1737  (1989); R. Alkofer and
 I. Zahed, Phys. Lett. B 238,  149 (1990).



 \bibitem{PRC1}
C.D. Roberts, R.T. Cahill, J. Praschifka, 
The effective action for the Goldstone modes in a
global colour symmetry model of QCD,
 Ann. of Phys. {\bf 188},   20 (1988).

\bibitem{holdom}
B. Holdom, Approaching low-energy QCD with a gauged,
nonlocal, constituent-quark model,  Phys. Rev. D 45,  2534 (1992).

\bibitem{ERV} D. Ebert, H. Reinhardt, M.K.Volkov, 
 Effective hadron theory of QCD,
Pr. Part. Nucl. Phys.
{\bf 33},  1 (1994).


\bibitem{lowdon}
P. Lowdon, 
Nonperturbative structure 
of the photon and gluon propagators,
Phys. Rev. D96, 065013 (2017).

\bibitem{cornwall}
J. M. Cornwall,  Entropy, confinement, and chiral symmetry
breaking, Phys. Rev. D 83,   076001 (2011).


\bibitem{higa} K. Higashijima,  
 Dynamical chiral-symmetry breaking,
Phys. Rev. D 29,  (1984) 1228.
V. A. Miransky,  Sov. J. Nucl. Phys. {\bf 38},  280 (1983).

\bibitem{aoki}
K.­I. Aoki, et al.,
Prog. Theor. Phys. {\bf 84}, 683  (1990).


\bibitem{SD-rainbow}
D. Binosi, L. Chang, J. Papavassiliou, C.D. Roberts,  
Bridging a gap between continuum-QCD and ab initiopredictions of hadron observables,
Phys. Lett. {\bf B 742},    183 (2015) and references therein.

\bibitem{gluon-prop-sde}
A. Bashir, et al.,
Collective Perspective on Advances in Dyson–Schwinger Equation QCD,
Commun. Theor. Phys. 58,79  (2012).
I.C. Cloet, C.D. Roberts, 
Explanation and prediction of observables using continuum strong QCD,
Prog. Part. Nucl. Phys. 77, 1 (2014).





\bibitem{EPJA2016}
 F.L. Braghin,  
Quark and pion effective couplings from polarization effects,
 Eur. Phys. Journ. {\bf A 52},   134 (2016).


\bibitem{PRD2019}
F.L. Braghin, 
Pion Constituent Quark Couplings strong form factors:
A dynamical approach, 
Phys. Rev. {\bf D 99},   014001 (2019).





\bibitem{EPJA2018} 
F.L. Braghin,
Low energy constituent quark and pion effective couplings in a weak external
magnetic field,
Eur. Phys. Journ. {\bf A 54}, 45 (2018) .

\bibitem{PRD2001}
F. L. Braghin,
Expanding nonhomogeneous configurations of the 
$\lambda \phi^4$
 model,
 Phys. Rev. D 64,    125001 (2001).
v
\bibitem{BFM}
L. F. Abbott, 
 Introduction to the background field method,
Acta Phys. Pol. B 13,  33 (1982).






\bibitem{mosel}
U. Mosel,  
Path Integrals in Field Theory,
An Introduction,  (2004)
Springer.




\bibitem{PRD2018}
F.L. Braghin,
Light vector and axial mesons effective couplings
to constituent quarks,
Phys. Rev. D97, 054025 (2018); 
Phys. Rev. D 101, 039901(E) (2020).



\bibitem{klimt-etal}
S. Klimt et al, Generalized SU(3) Nambu-Jona-Lasinio model (I),
Nucl. Phys. A516, 429 (1990).



\bibitem{eguichi-sugawara}
T. Eguchi, H. Sugawara, 
Extended model of elementary particles based on an analogy with superconductivity,
Phys. Rev. D10, 4257 (1974).
T. Eguchi,
New approach to collective phenomena in superconductivity models,
 ibid 14, 2755 (1976).


\bibitem{cutoff-IR}
D. Ebert, T. Feldmann, H. Reinhardt,
Extended NJL model for light and heavy mesons without q$\bar{q}$
thresholds,
Phys. Lett.
B 388,  154 (1996).



\bibitem{SBK}
F. E. Serna, B. El-Bennich,  G. Krein,
Charmed mesons with a symmetry-preserving contact interaction,
Phys. Rev. D96, 014013 (2017).

\bibitem{kohyama-etal}
H. Kohyama, D. Kimura, T. Inagaki,
Parameter fitting in three-flavor Nambu-Jona-Lasinio model with 
various regularizations,
Nucl. Phys. B906, 524 (2016).

{



\bibitem{alkofer-etal}
H. Reinhardt, R. Alkofer, 
Instanton-induced flavour mixing in mesons, 
Phys. Lett. B 207, 482 (1988).


\bibitem{osipov-etal}
A.A. Osipov, B. Hiller, Eur. Phys. J. C 35, 223 (2004).
3. A.A. Osipov, B. Hiller, J. Moreira, A.H. Blin, Eur. Phys. J. C 46, 225 (2006).
4. A.A. Osipov, B. Hiller, A.H. Blin, Eur. Phys. J. A 49, 14 (2013).

 
\bibitem{jamin}
M. Jamin,
Flavour-symmetry breaking of the quark
condensate and chiral corrections to the
Gell-Mann-Oakes-Renner relation
Phys.Lett. B538, 71  (2002).

\bibitem{davies-etal}
C.T.H. Davies {\it et al},
Determination of the quark condensate from heavy-light current-current correlators
in full lattice QCD,
Phys. Rev. D 100, 034506 (2019).

\bibitem{flag}
S. Aoki, FLAG, Review of lattice results concerning low energy particle physics,
Eur. Phys. Journ. C  80,  113 (2020).


}



\end{thebibliography}
\end{document}